\documentclass[12pt,preprint]{aastex}

\shorttitle{Cosmic ray acceleration at relativistic shocks}
\shortauthors{Niemiec, Ostrowski}
\usepackage{natbib}

\begin{document}

\title{COSMIC RAY ACCELERATION AT RELATIVISTIC SHOCK WAVES \\
 WITH A `REALISTIC' MAGNETIC FIELD STRUCTURE}

\author{Jacek Niemiec}
\affil{Instytut Fizyki J\c{a}drowej Polskiej Akademii Nauk,\\
ul. Radzikowskiego 152, 31-342 Krak\'{o}w, Poland}
\email{Jacek.Niemiec@ifj.edu.pl}
\and
\author{Micha\l{} Ostrowski}
\affil{Obserwatorium Astronomiczne, Uniwersytet Jagiello\'{n}ski, \\
ul. Orla 171, 30-244 Krak\'{o}w, Poland}

\begin{abstract}
The process of cosmic-ray first-order Fermi acceleration at relativistic shock 
waves is studied with the method of Monte Carlo simulations. The simulations are
based on numerical integration of particle equations of motion in a
turbulent magnetic field near the shock. In comparison to  earlier studies, 
a few 
``realistic'' features of the magnetic field structure are included. The upstream
field consists of a mean field component inclined at some
angle to the shock normal with finite-amplitude sinusoidal perturbations imposed
upon it. The perturbations are assumed to be static in the local plasma 
rest frame. Their flat or Kolmogorov spectra are constructed  
with randomly drawn wavevectors from a wide range $(k_{min}, k_{max})$.
The downstream  field structure is derived from the upstream one as
compressed at the shock. We present and discuss particle spectra 
and angular distributions obtained at mildly relativistic sub- and superluminal
shocks. 
We show that particle spectra diverge from a simple power law; the exact
shape of the spectrum depends on both the amplitude of the magnetic field
perturbations and the wave power spectrum considered. Features such as spectrum
hardening before the cut-off at oblique subluminal shocks and formation of
power-law tails at superluminal ones are presented and discussed.
The simulations have also been 
performed for parallel shock waves. The presence of
finite-amplitude magnetic field perturbations leads to the formation of locally 
oblique field configurations at the shock and the respective magnetic field
compressions. This results in the modification of the particle acceleration
process, introducing some features present in oblique shocks, e.g.,
particle reflections from the shock. For the first time, we demonstrate for
parallel
shocks a (nonmonotonic) variation of the accelerated particle spectral index
with the turbulence amplitude. At the end, a few astrophysical consequences of 
the  results we obtained are mentioned.
\end{abstract}

\keywords{acceleration of particles, cosmic rays, gamma rays:bursts, methods:
numerical, relativity, shock waves}

\section{INTRODUCTION}
Relativistic shock waves are postulated to be responsible for a number of
phenomena of high-energy astrophysics. Many sources of high-energy radiation
appear at the same time to involve highly relativistic plasma flows, indicating
that the relativistic shocks are the acceleration sites of energetic particles 
producing high-energy emission. The observed examples of such flows/shocks  
can be hot spots in radio galaxies, parsec-scale 
jets in blazars, jets in galactic ``microquasars'', gamma-ray burst sources,
and pulsar winds. The understanding of particle 
acceleration processes at relativistic shocks is essential in the study of these 
objects.

At relativistic shock waves the bulk flow velocity is comparable to 
particle velocity. This leads to anisotropy of particle
angular distribution, which can substantially influence the process of particle
acceleration. 
The first consistent method to tackle the problem for a parallel
shock wave, where the mean magnetic field is parallel to the shock velocity, 
was proposed by \citet{kir87a}. 
Generalizing the diffusive approach by explicit treatment of the 
particle distribution function anisotropy, they constructed stationary solutions 
of the pitch-angle diffusion equation on both sides of the shock. The 
requirement of distribution function continuity at the shock  allowed them to
match the upstream and downstream solutions by taking into account a
sufficient number of high-order terms in particle anisotropy. The solution was
obtained from
numerical fitting of the respective coefficients. This semi-analytic procedure
yielded the power-law index of the resulting spectrum, as well as the anisotropic 
particle angular 
distribution at the shock. The spectral indices for the phase space distribution
function\footnote{The energy spectral index of the particle distribution
function $N(E)\propto E^{-\sigma}$, often used in the gamma-ray burst studies, 
is $\sigma =\alpha -2$.}
derived with this method for 
relativistic shocks --- with the compression ratio $R=4$ --- were only slightly 
different from
the $\alpha = 4$ value obtained in the nonrelativistic case. These findings were
later confirmed by \citet{kir87b} with the method of particle Monte
Carlo simulations (see also Ellison et al. 1990).

An extension of the above approach was
proposed by \citet{hea88}, who modified the method to treat more
general conditions at the shock. They investigated the fluid dynamics of
relativistic shock waves and applied the results to calculate 
spectral indices of particles accelerated at such shocks. For a parallel shock
wave propagating into cold electron-proton or electron-positron plasma, 
they found that the particle spectral index depends on the form of the wave 
power spectrum, in contrast to the nonrelativistic case. 
The spectral indices  obtained were close to the respective values given by the
nonrelativistic formula, $\alpha = 3R/(R-1)$.
\citet{kir89} considered the acceleration processes in shocks with
magnetic fields oblique to the shock normal. 
For the subluminal shock
configurations studied they assumed magnetic moment
conservation for particles interacting with the shock. This assumption 
restricted the validity of their considerations to the case of a weakly 
perturbed magnetic field. 
The important result of this work was that oblique shocks lead to flatter 
spectra than parallel ones, a spectrum could be even as flat as 
$\alpha\approx 3$ in cases where the shock velocity along magnetic field lines 
is close to the speed of light. This feature arises because of effective multiple 
reflections of anisotropically distributed upstream
particles from the compressed downstream magnetic field. However, as pointed
out by \citet{beg90}, most field configurations in relativistic shocks 
lead to superluminal conditions for the acceleration process. Therefore,
unless there is strong particle scattering across field lines, particles are
not able to cross the shock front repeatedly, and diffusive acceleration
is inefficient.
Instead, particle energy gains are due to the shock-drift acceleration 
process,
which involves a single shock transmission of upstream particles downstream of
the shock (Webb et al. 1983; Drury 1983).  

The above approximations rely on an assumption of a weakly perturbed magnetic
field, where cross-field diffusion does not play a significant role.
However, in
astrophysical shocks one may expect large-amplitude MHD waves to occur.
In such conditions, scattering of particles is not exclusively related to
resonant interactions with the magnetic field fluctuations, as in the
small-amplitude wave limit. In this case, the
Fokker-Planck equation does not properly describe the particle transport, nor 
does the 
particle magnetic moment conservation hold for oblique shocks
\citep{dec85,ost91}. Thus, the particle acceleration
process has to be investigated with the help of numerical methods.  

The role of finite-amplitude magnetic field perturbations in forming a
particle spectrum was investigated by a number of authors using Monte Carlo 
particle simulations \citep[e.g.,][]{ell90,ost91,ost93,bal92,nai95,bed96,bed98}. 
The main result of these
considerations was the finding of a direct dependence of the derived particle 
spectra on the conditions
near the shock. The power-law spectra can  be either very steep or very flat 
for different mean magnetic field inclinations to the shock normal and different
amplitudes of perturbations, where changes of the particle spectral index can be
non-monotonic with an increasing field perturbations amplitude
\citep{ost91,ost93}.

The process of the first-order Fermi acceleration at ultrarelativistic shocks
($\gamma\gg 1$) has been studied in the last few years by a number of authors
\citep{bed98,gal99,gak99,ach01,kir00,lem03}. All these approaches
yield consistent estimates of the asymptotic ($\gamma\rightarrow\infty$)
spectral index $\alpha\approx 4.2-4.3$ in the limit of highly turbulent
conditions near
the shock \citep[see][]{ostb02}. In conditions with 
medium-amplitude magnetic field perturbations, the particle spectra generated at
ultrarelativistic shocks can be much steeper than those
obtained in the asymptotic limit, as seen in the simulations of \citet{bed98}.

Numerical investigations of the first-order Fermi process are based on 
simulations of particle motion in a turbulent magnetic field near the shock
front. One possible
method is to use the pitch-angle diffusion model to study particle
transport in a way similar to the quasi-linear analytic approximation. 
Such an approach was first presented by \citet{kir87b}, 
and essentially 
the same technique was independently elaborated for oblique shocks by 
\citet{ost91} \citep[see also][]{bed96,bed98}. In this approach, a particle 
is assumed to move in a regular magnetic field along its
undisturbed ``adiabatic'' trajectory, and the influence of the turbulent field on
the orbit is described by discrete, uncorrelated, small-amplitude particle 
pitch-angle 
perturbations in finite time intervals. Depending on the model parameters ---
the mean time between successive scattering acts and the scattering amplitude ---
one is able to simulate situations with different magnetic field configurations
and different turbulence amplitudes. 

The other and perhaps more basic method of
numerical simulations is to specify the entire magnetic field in the space near
the shock and to integrate particle equations of motion in such a ``realistic'' 
field. However, this method
is much more difficult to implement because precise numerical calculations
of particle trajectories in a wide wavevector range spectrum  require usually 
excessive simulation times. Because of this fact, the two early papers adopting this 
technique for relativistic shock waves had to apply very simple models of a 
turbulent magnetic field.
\citet{bal92} considered a highly disordered magnetic field with
vanishing regular component, equivalent to discussing a parallel highly
turbulent shock. The field was described in  Fourier space by the 
Kolmogorov power-law
spectrum in a very limited wavevector range, $k_{max}/k_{min}\approx 20$.
Because of this restricted dynamical range for the magnetic field perturbations, 
the resulting particle spectra were formed in a limited energy range near the
injection energy. 
A different approach was presented by \citet{ost93}. In his work, the magnetic
field consisted of the regular component and the
perturbations, described as the superposition of a few sinusoidal
static waves of finite amplitude, oriented randomly with respect to the mean
field. For each particle of a given energy there were 
three wavevector ranges, containing short, resonance, and long waves with the
flat power spectrum. When in the course of the acceleration process a particle
energy increased, the respective short waves were replaced by the existing 
previously resonance wavevectors, the resonance waves by the long ones, 
and a new set of waves
was selected for the long waves. 
With this procedure, one was able to study particle acceleration to very high 
energies,
which allowed for an accurate determination of the spectral indices in an 
energy range not influenced by the initial conditions. The spectra obtained by
Ostrowski for parallel shocks were 
systematically flatter than those calculated by Ballard \& Heavens (1992)
\citep[see discussion in][]{ost93}.

The purpose of our present work is to simulate the first-order Fermi
acceleration process at (mildly) relativistic shock waves, propagating  in  
more realistically modeled perturbed magnetic fields. This involves the use of 
a wide wavevector
range turbulence with a power-law spectrum and continuity of the magnetic 
field across the shock front, according to the respective matching conditions.
Particle trajectories are calculated by integrating their equations of motion in
such fields. Both sub- and superluminal mean magnetic field configurations are 
studied, as well as the particular case of parallel shocks.
Angular distributions and wide-energy particle spectra are derived
in a range of model parameters. We show that the particle spectra diverge from a
simple power law in the full energy range considered, yielding in some cases
interesting spectral features. A number of previously obtained results are also
reproduced.
The simulation method and the magnetic field structure assumed are
described below, in \S 2. The results are presented in \S 3 and 
finally summarized and discussed in \S 4. 

Below, the speed of light $c$ is used as the velocity unit. All calculations are
performed in the respective local plasma (upstream or downstream) rest frames.
The upstream and downstream quantities are labeled with 
the indices 1 and 2, respectively. We consider ultrarelativistic particles with $p=E$.
In the units we use in our simulations,
a particle of unit energy moving in a uniform mean upstream magnetic
field $B_{0,1}$ has the unit maximum (for $p_{\perp}=E$) gyroradius 
$r_g(E=1)=1$ and the respective resonance wavevector $k_{res}(E=1) = 2\pi$.

\section{SIMULATIONS}
In our simulations, trajectories of ultrarelativistic test particles are derived
by integrating their equations of motion in a perturbed magnetic field.
A relativistic shock wave is considered to be a planar discontinuity, propagating
in a rarefied (collisionless) electron-proton plasma with a turbulent 
magnetic field frozen in it. 
Upstream of the shock, the field is assumed to consist of the uniform component
 {\bf\em B}$_{0,1}$, inclined at some angle $\psi_1$ to 
the shock normal, 
with finite-amplitude perturbations imposed upon it. The irregular
component has either a flat $(F(k)\sim k^{-1})$ or a Kolmogorov 
$(F(k)\sim k^{-5/3})$ wave power spectrum in the (wide) wavevector range 
$(k_{min}, k_{max})$. The perturbations are assumed to be static in the local 
plasma rest
frame, both upstream and downstream of the shock. Thus, the possibility of 
second-order Fermi acceleration is excluded from our considerations. The shock 
propagates
with velocity $u_1$ with respect to the upstream plasma. The magnetic field 
is assumed to be dynamically unimportant and its downstream structure, together
with the downstream flow velocity $u_2$, are obtained from 
hydrodynamic jump conditions for shocks propagating in the cold electron-proton 
plasma. Derivation of the shock compression ratio,  
defined  in the shock rest frame as $R = u_1/u_2$, is
based on the approximate formulae derived by Heavens \& Drury (1988). 
For the case of the shock propagating with  velocity $u_1=0.5c$ studied in
this paper, 
these formulae give $R\simeq 5.11$. For the shock
velocity $u_1=0.9c$,  the shock compression ratio is $R\simeq 3.92$ and for 
$u_1=0.98c$ $R\simeq 3.29$.
The acceleration process is considered in the particle energy range
where radiative (and other) losses can be neglected. 

The model of the medium near the shock used in the paper is constructed 
by four
elements: the turbulent magnetic fields upstream and downstream of the shock and
the two free escape boundaries, located far upstream and far downstream from the
shock front. The magnetic field structure is described below in \S 2.1, and
the cause for introducing the boundaries is explained in \S 2.2. 

\subsection{Magnetic Field Structure}
The `1turbulent'' ($\equiv$ perturbed) magnetic field component upstream of the 
shock is modeled as a superposition of
sinusoidal static waves of finite amplitude \citep[see][]{ost93}.
In our simulations, in a magnetic field-related primed coordinate 
system,\footnote{In the upstream plasma rest frame  
the $x$-axis of the (unprimed) Cartesian coordinate system $(x, y, z)$ is
perpendicular to the shock surface. The shock wave moves in the negative 
$x$-direction and the regular magnetic field  lies in the $x-y$ plane. 
The primed system $(x', y', z')$ is obtained from the unprimed one by its
rotation about the $z$-axis by an angle $\psi_1$, so that the $x'$-axis is 
directed along {\bf\em B}$_{0,1}$.} 
they take the form:
\begin{equation}
\delta B_{x'} = \sum_{l=1}^{294} \delta B_{x'l} \,\sin (k_{x'y'}^{l} y' + 
k_{x'z'}^{l} z'), 
\end{equation}
where ${k_{x'y'}^{l}}^{2} + {k_{x'z'}^{l}}^{2}  = {k_{x'}^{l}}^2$ (see below),
and analogously for $\delta B_{y'}$ and $\delta B_{z'}$ components. 
Such a form of
$\delta${\bf\em B} ensures that $\nabla\cdot${\bf\em B}$=0$.
The index
$l$ enumerates the wavevector range from which the wave 
vectors $k_{i}^{l} \: (i = x',y',z')$ are randomly drawn. 
The components $k_{x'y'}^{l}$, $k_{x'z'}^{l}$ of the wavevectors 
$k_{x'}^{l}$ are selected by choosing a random phase
angle $\phi _{x'}^{l}$, so that $k_{x'y'}^{l}=k_{x'}^{l}\cos\phi_{x'}^{l}$ and 
$k_{x'z'}^{l}=k_{x'}^{l}\sin\phi_{x'}^{l}$, and analogously for other 
components. The wavevectors span the
range $(k_{min}, k_{max})$, where $k_{min} = 0.0001$ and $k_{max} = 10$.
The selected form of the perturbations and the method of $k_{i}^{l}$ vector
components drawing both produce isotropic turbulence.
The number of 294 wavevector ranges used in the simulations has been selected 
on the basis of a series of numerical tests. 
It has been checked that this number is sufficient for the perturbed 
magnetic field to diffusively scatter particles. Using a larger
number of waves would result in longer simulation times. On the
other hand, taking a substantially smaller number of waves could lead to 
resonance (nonstochastic) features in the particle trajectories 
(see Karimabadi et al. 1992). 
 
The wave amplitudes $\delta B_{il}$ are selected at random to satisfy 
${\delta B_{l}}^2 = \sum_{i} {\delta B_{il}}^2$, and the amplitudes 
$\delta B_{l}$ are chosen to reproduce the turbulence power spectrum 
assumed.
In the wavevector range considered, this spectrum can be written
\begin{equation} 
\delta B_{l}(k) = \delta B_{l}(k_{min}) \left( \frac{k}{k_{min}}
\right)^{(1-q)/2},
\end{equation}
where $q$ is the wave spectral index; $q=1$ corresponds to the flat spectrum and
$q=5/3$ to the Kolmogorov one. The constant $\delta B_{l}(k_{min})$ is scaled 
to match the model parameter
$\delta B \equiv [\sum_{l} {\delta B_{l}}^2]^{1/2}$. 
Its ratio to the upstream mean magnetic field, $\delta B / B_{0,1}$, is our
measure of the field perturbations amplitude.

The Kolmogorov wave power spectrum  $F(k)\sim k^{-5/3}$ 
($\delta B^2=\int F(k)dk$) seems to be an appropriate (although still
approximate) choice for magnetic field 
turbulence description in real astrophysical plasmas. Such a spectrum is 
measured
in the solar wind \citep{lea98}. Interstellar scintillation observations
indicate that the electron density spectrum follows the Kolmogorov slope 
\citep{lee76,lag83,cho01,sti00} (see, however, Boldyrev et al.
2002) and the same
character could probably be ascribed to the MHD turbulence power spectrum
(Cho 2001; Cho et al. 2003). Such a form of the spectrum of 
perturbations was also
often used in modeling of acceleration processes at shocks. 
The flat turbulence power spectrum,
$F(k)\sim k^{-1}$, is considered here in order to check how the accelerated
particle spectra depend on the value of the wave spectral index, and also to
compare our results to the earlier studies. The simulations are performed for 
the case of a weakly ($\delta B / B_{0,1}\ll 1$) and a highly 
($\delta B / B_{0,1}\ga 1$) perturbed magnetic field. The finite-amplitude
magnetic field perturbations can exist in the plasma where the shock wave
propagates, as observed, e.g., for heliospheric plasma, where 
$\delta B / B_{0,1}\sim 1$ \citep[see][]{ell90}. Calculations for 
nonrelativistic shocks indicate that the
turbulence can be also produced upstream of the shock by  nonlinear 
interactions between MHD waves generated by accelerated particles
\citep{dru83,lag83,luc00}. In such a case, the
resulting form of the turbulence spectrum is not satisfactorily known.

The downstream magnetic field structure is obtained from the upstream one as
compressed at the shock. According to the gas compression conditions,  
any given point $(x_2, y, z)$ in the downstream plasma was at the position 
$(x_1=r x_2, y, z)$ in the upstream rest frame before the shock passage.
One should note that the upstream and the downstream magnetic fields are 
compared in two time instants, before and after the shock passage, respectively.
The compression factor $r=R\gamma_1/\gamma_2$, where $R$ is the shock compression
ratio in the shock-normal rest frame and 
$\gamma_i=1/\sqrt{1-u_i^2} \; (i = 1,2)$. 
Thus, to derive the downstream
field components at the point $(x_2, y, z)$ we calculate the unshocked magnetic 
field at the point $(x_1, y, z)$ and compress the tangential components of the
magnetic field by the factor $r$. The field component parallel to the shock normal
remains unchanged. Note that the magnetic field derived in such a way is 
continuous across the shock. In our approach, the shock front itself does not 
produce any additional
fluctuations of the field. However, as pointed out by Medvedev 
\& Loeb (1999) \citep[see also][]{nis03,fre03a,fre03b}, the relativistic 
two-stream
instability can generate small-scale strong magnetic field perturbations in the
vicinity of the shock. Thus, our modeling can be considered valid only for 
high-energy
particles, with gyroradii much larger than the spatial scale of
inhomogeneities thus generated. A study of the influence these small-scale
perturbations have on the particle spectra formation is planned in future work.

\subsection{Simulation Method}
The particle equations of motion are integrated in the local 
(upstream or downstream) 
plasma rest frame, where the electric field vanishes. At the beginning of the
simulation run,  $N$ monoenergetic particles (usually $N=100$) are
injected at the shock front with their momentum vectors directed into the upstream 
region. Particle injection energy measured in the downstream rest frame is 
$E_0=0.1$ 
and their resonance wavevectors $k_{res}(E_0) \gg k_{max}$
\footnote{For superluminal shocks the seed particle injection procedure is 
modified. In this case, the majority of injected particles escape downstream
after the first interaction with the shock and only a few particles are still able
to take part in the acceleration process. The resulting spectra are then 
determined with poor accuracy. To overcome this problem, we increase the number
of injected particles in such a way as to select $N$ of those that, after
transmission downstream, succeed in reaching the shock front again. 
We have checked that this method gives the same results when applied to
subluminal shocks;
the only difference appears at the energy range close to $E_0$, which is already 
influenced by the initial conditions.}.
After the Lorentz transformation to the upstream plasma rest frame,
the trajectory of each particle is
followed until it crosses the shock surface or reaches an introduced free escape
boundary, located far 
upstream from the shock. In the latter case, the particle is assumed to
escape from the acceleration region. The boundary is required to treat 
high-energy particles, with resonance wavevectors much smaller 
than $k_{min}$, but also the low-energy particles, with 
$k_{res}\gg k_{max}$ 
\footnote{The upstream escape boundary for 
low-energy particles with $k_{res}\gg k_{max}$ is introduced only in the case
of oblique subluminal
and parallel shock waves, propagating with  velocity $u_1=0.5c$ in the weakly
perturbed magnetic field $(\delta B/B_{0,1}=0.3)$. In these cases, the upstream
boundary is located at several tens of thousand of $r_g$ ahead of the shock 
front. 
The escape boundary for energetic particles ($k_{res}\ll k_{min}$) is located at 
200 $r_g(E)$ ahead of the shock in the case of oblique subluminal shocks, 
whereas for parallel shocks it is
placed at $2000 r_g$ upstream of the shock. The number of low-energy particles 
escaping upstream is small. However, for the highest energy particles, with 
$k_{res}(E)\ll k_{min}$, the escape becomes efficient, leading to the formation
of the cutoff in particle spectra.}.
Such particles are only weakly scattered by the turbulent field and 
can propagate very far upstream from the shock.
The upstream particles that cross the shock
front are Lorentz-transformed into the downstream plasma rest frame. 
At this side of the shock a free escape boundary is located ``far downstream'' 
from the shock at 
$x_{max}(E) = X_{max}\, r_g(E)$ behind it. The selected value of the coefficient 
$X_{max}\in (5,150)$
depends on simulation parameters and is specified by numerical tests, separately 
for each shock configuration.
In the conditions considered by us, very few particles are able to diffuse from 
such a distance back to the shock.  The trajectory of a particle downstream of 
the shock 
is integrated until it crosses the escape boundary or reaches the shock front.
After performing this procedure for all $N$ particles, the first simulation 
cycle 
is finished. Then the splitting procedure is performed (see below) to replace
all escaped particles with the ones still active in the acceleration process, 
with the respective division of the particle weights.
Subsequent cycles calculations are performed analogously. 
The computations are finished when either more than
90\% of particles escape through the introduced boundaries in an individual 
simulation
cycle or all particles reach the upper energy boundary $E_{max}=10^8$ or the 
lower limit for a particle weight $w_{min}$ (see below), which is usually set
to $10^{-6}w_0$. We have checked that allowing for further calculations does not 
noticeably influence the spectrum formed.   

We use the method of trajectory splitting to derive spectra in a wide 
energy range. This method is a standard tool of Monte Carlo particle simulations
(e.g., Hammersley \& Handscomb 1965; see also Kirk \& Schneider 1987b; 
Ostrowski 1991, 1993). 
At the beginning of the simulations, the same initial weight $w_0$ is ascribed to
each particle. During the calculations, some particles are lost by escaping 
through the boundaries. To keep the total number of particles active in the
simulations constant, for each lost particle we
duplicate (or ``multiplicate'') one of those that crossed the shock upstream, with 
the weight of the original particle divided at equal parts between the
``daughter'' particles. Then we weakly perturb the motion 
of these particles to create different trajectories out of the split one.
The perturbations are performed by a small change in particle positions in the
shock plane, $\Delta y, \Delta z \sim 0.0002 \lambda_{min}(E)$, where 
$\lambda_{min}(E)$ is the smallest scale of the resonant range of the field 
perturbations, defined in \S 2.3. For the splitting procedure we 
take particles with the highest weights and smallest energies available. 
With this method, the effective number of particles in the simulation is much
higher than the injected number, and a particle spectrum is determined with
approximately the same accuracy at all energies.

A spectrum of accelerated particles and their angular distribution are 
measured at the shock, in the shock rest frame. Every time a particle crosses
the shock front, a value $w/(|v_x|+0.005)$ is added to the respective logarithmic
energy bin and to the respective $\cos\theta$ bin. Here $\theta$ is the angle between
the particle momentum and the shock normal (note that $\theta$ is not the 
pitch angle measured with respect to the mean magnetic field), $v_x$ is a normal
component of the particle velocity, and $w$ is 
its weight. The weighting 
factor $(|v_x|+0.005)$ is needed to transform the particle flux into the
particle density (see Kirk \& Schneider 1987b)
and the additional term 0.005 ensures proper behavior near $v_x=0$.
The spectra and angular distributions presented are averaged ones over a few 
(usually 10) different sets of $N$ particles and realizations of the perturbed 
magnetic field.

\subsection{Derivation of Particle Trajectories}
The particle orbits are calculated using the Runge-Kutta fifth-order method with
the adaptive step size control (routine RKQS in Press et al. 1992). 
This routine allows for an accurate derivation of particle trajectories in a
turbulent magnetic field with the optimized simulation time 
(see also Appendix B). 
However, the procedure becomes very time consuming for higher energy particles 
because exact derivation of a trajectory with short-wave perturbations requires 
extremely short time steps (in units of $r_g/c$) to be used. At the same time, the
considered short waves, with $\lambda \ll r_g$, only weakly influence
trajectories of such particles. 
Therefore, in our modeling the motion of a higher energy particle is derived 
within the proposed hybrid approach. It involves exact integration of 
particle trajectories in the turbulent field including long and resonance waves
($\lambda > \lambda_{min}=0.05 r_{g}(B=B_{0,1})$),\footnote{The short waves
downstream of the shock are defined by a particle gyroradius in the upstream
mean magnetic field $B_{0,1}$. However, this assumption does not influences the
particle spectra formed, as discussed in Appendix B.} and the short-wave
influence on the trajectory is included as a respective small-amplitude momentum
scattering term. A scattering probability distribution is determined by
additional simulations, involving the full set of the short waves considered
(see Appendix A). Example particle trajectories are shown in Figure 1.

\begin{figure}[t]
\plotone{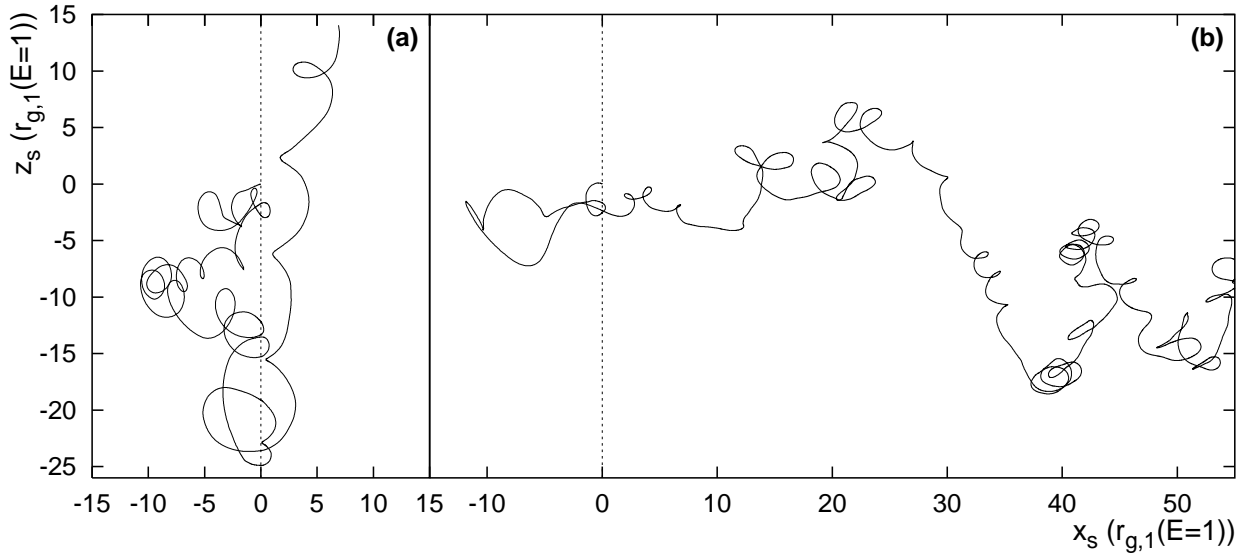}
\caption{\small Example particle trajectories projected onto $x_s-z_s$ plane in the
shock normal rest frame. The trajectories were calculated for particles of
initial energy $E_1=1$, injected at the shock with different initial conditions
(a momentum inclination to the shock normal, the magnetic field strength and 
inclination
at the injection point) and shock parameters: $u_1=0.5c$, $\psi_1=45^o$ and 
$\delta B / B_{0,1}=1.0$. The shock front is shown by a dotted line and the
trajectories originate at $(0,0)$. The effects of the turbulent magnetic field
structure, including the presence of long-wave perturbations, are visible in the
figure. The particle in (a) crosses the shock front several times and is
accelerated up to $E_2\approx 11.8$. The particle in (b) is accelerated up
to $E_2\approx 3.3$ and then escapes far downstream from the shock.
\label{fig1}}
\end{figure}

\subsection{Tests}
A number of tests have been performed on the simulation code. Some of them have been
mentioned in the preceeding subsections; the others are described in
detail in Appendix B.

\section{RESULTS}
Our Monte Carlo simulations of the first-order Fermi acceleration process have 
been performed
in order to study the role of the applied magnetic field structures  
for the test particle acceleration at relativistic shocks. Because of
the limited dynamic range of field perturbations, scattering conditions 
vary with increasing particle energies. We study how this influences the
particle spectrum formation and how the spectra depend on the shock velocity
and on the parameters of the magnetic field: the mean field 
inclination and the spectrum of finite-amplitude perturbations. 
In the case of oblique subluminal shocks, the accelerated 
particle spectral index can vary with the turbulence amplitude. At superluminal
shocks, the presence of finite-amplitude perturbations near the shock plays an 
essential role in particle acceleration. In 
weakly scattering conditions, the first-order Fermi process is inefficient in 
such shocks and particles can gain energies only from the one-way compression at 
the shock (Begelman \& Kirk 1990). Formation of power-law spectra is 
possible for
larger amplitudes of the magnetic field perturbations. In particular, we discuss
the role of long-wave perturbations, which lead to the formation of locally 
oblique subluminal configurations at the shock. In the conditions provided, 
the energy gains of particles interacting with the shock can increase and 
particle reflections from the respective regions of the compressed field 
downstream of the shock may occur. The importance of this factor on  
particle spectra formation is investigated  for parallel shocks as well. 

Accelerated particle spectra formed at
oblique shock waves are presented in Figure 2 for the case of subluminal
magnetic field configuration, and in Figures 5 and 9 for superluminal shocks.
Particle spectra derived for parallel shocks are shown in Figure 10.
The particle spectra were calculated for three different upstream 
magnetic field perturbation amplitudes $\delta B/ B_{0,1}=0.3, 1.0$ and $3.0$, 
referred to below as the small, medium and  large (high) amplitude
perturbations, respectively. The amplitudes $\delta B/ B_{0,1}$ 
are indicated in the figures near the respective curves. Spectra for different 
$\delta B/ B_{0,1}$ are not normalized; they have vertical shifts for
presentation clarity. 
Linear fits to the power-law parts of the spectra are also presented in the
figures, and values of the spectral indices $\alpha$ are given. Accuracies of the 
presented fits 
are usually better than 0.02 in $\alpha$ for  
subluminal and parallel shocks, with some exceptions discussed in the following.
For superluminal shocks, the evaluated statistical errors 
$\Delta\alpha\approx 0.03-0.06$. One should note that the power-law 
character of some particle spectra is only approximate because of the existing 
spectrum curvature (see, e.g., Figures 5d, 5f and 9b) and thus that the value of the 
spectral index derived depends also directly on an energy range 
chosen for the fit. The narrow energy range values of $\alpha$, indicated in 
Figures 5 and 9, are
shown for comparison to the spectral indices of the other spectra presented.

Particles in the energy range indicated in the figures by arrows above the
horizontal energy axis can 
effectively interact with the magnetic field inhomogeneities because of the
 resonant condition satisfied,
$k_{min}\la k_{res}\la k_{max}$ ($k_{res}\equiv 2\pi/r_g$). The limits
are not precise; they are calculated for the upstream magnetic field strength
$B_{0,1}=1$ and particle transverse momenta $p_{\perp}=p$. The
particle energies considered in the simulations can be scaled to various shock
conditions by the respective identifications of $B_{0,1}$, $\psi_1$, $\delta B$,
$k_{min}$, and $k_{max}$.

\begin{figure}[t]
\plotone{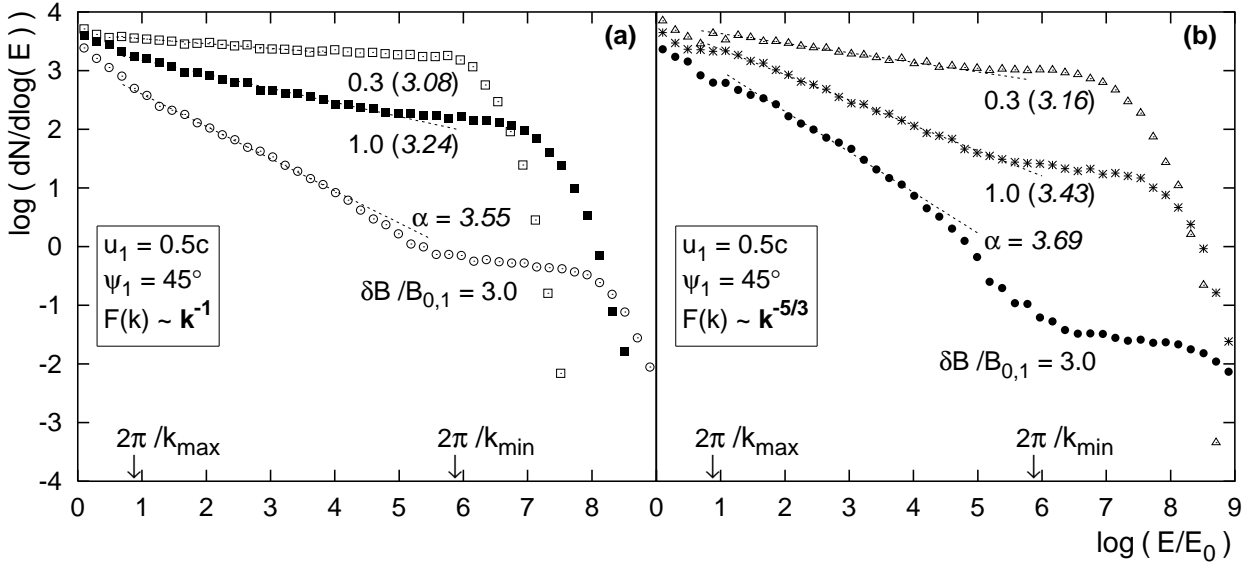}
\caption{\small Particle spectra at the subluminal shock wave
($u_1=0.5c$, $\psi_1=45^o$ and $u_{B,1}\simeq 0.71c$)  
for (a) the flat and (b) the Kolmogorov wave spectrum of magnetic field 
perturbations. Any individual point in the spectrum represents a particle number
(weight) $dN$
recorded per a logarithmic energy bin. The upstream perturbation amplitude 
$\delta B / B_{0,1}$ is given near the respective results. Linear fits to the 
spectra are also presented and values of the spectral indices $\alpha$ are given
in italic. The spectra have vertical shifts for clarity. Particles with energies 
in the range ($2\pi/k_{max}, 2\pi/k_{min}$) indicated by arrows can effectively
interact with the magnetic 
field inhomogeneities. \label{fig2}}
\end{figure}

\subsection{Oblique Subluminal Shock Waves}
To illustrate the characteristic features of 
particle acceleration processes at relativistic oblique subluminal 
($u_1/\cos \psi_1 < c$) shocks, we have studied  the shock wave 
propagating with velocity $u_1 = 0.5c$ ($\gamma_1\simeq 1.2$) into the 
upstream plasma with the mean magnetic field inclined  to the shock normal at 
$\psi_1 = 45^o$. The shock
velocity along the mean magnetic field is then $u_{B,1}\simeq 0.71c$. 
Figure 2 shows 
accelerated particle spectra, derived in the shock rest frame, for
the flat and the Kolmogorov wave power spectrum.   
One may note that the particle spectra diverge from a power-law in the full 
energy range. Usually a harder spectral component appears at the highest energies, 
followed by the spectrum cutoff. The shape of the particle spectrum depends on 
both the amplitude of the magnetic field perturbations and the spectral index 
of the wave power spectrum.

The power-law part of the particle spectrum steepens with the increasing amplitude of
the field perturbations for the three cases considered. For the flat turbulence 
spectrum, the particle spectrum is very flat $(\alpha \approx 3.08)$ for 
the weakly perturbed field $\delta B / B_{0,1} = 0.3$. For larger perturbation 
amplitudes we have $\alpha \approx 3.24$ and $\alpha \approx 3.55$ for 
$\delta B/ B_{0,1} = 1.0$ and  $3.0$, respectively. 
The spectral indices obtained are consistent with previous
numerical calculations of Ostrowski (1991, 1993) and, in the limit of small 
perturbations, with the analytic results obtained  by Kirk \& Heavens (1989).

Variations of the accelerated particle spectral index with the turbulence 
amplitude, for a given mean field inclination, was discussed by 
Ostrowski (1991). As already mentioned in \S 1, in a weakly 
perturbed magnetic field there is a high probability of reflection of upstream 
particles from the compressed field behind the shock. The repeating particle
reflections lead to the formation of a
very flat energy spectrum. When the perturbation amplitude increases, the
reflection probability is diminished and the mean energy gains of particles
interacting with the shock decrease. At the same time, the escape probability for
downstream particles decreases, since the field perturbations enable particles to 
diffuse across the field lines. However, this reduction of the escape probability 
is slow, so that increased upstream-downstream transmission probability leads 
to a net more
efficient particle escape and the resulting steepening of the spectrum.

The differences in the particle spectra obtained for the Kolmogorov and the flat
wave power spectrum are clearly visible from comparison of Figures 2a and 2b.
In particular, for a given $\delta B / B_{0,1}$ the power-law part of the 
spectrum is steeper for the Kolmogorov wave spectrum,
and the final hardenings of the 
spectra are more pronounced in this case. 

The non--power-law character of the obtained particle spectra results from the 
limited dynamic range of the magnetic field perturbations. In the energy range
where an approximate power-law spectrum forms, particles are effectively 
scattered by the resonant magnetic field inhomogeneities. 
The character of the spectrum changes at high particle energies, for which 
resonance wavevectors $k_{res}\la k_{min}$. These particles interact with
the magnetic field perturbations being the short-wave turbulence and are only 
weakly scattered. Thus, analogously to weakly scattering conditions, 
the anisotropically distributed upstream particles can be 
effectively reflected from the region of the compressed magnetic field 
downstream of 
the shock, leading to the spectrum flattening. The effectiveness of reflections 
and the resulting modification of the spectrum 
depend on the amplitude and the spectrum of  field perturbations, as can be seen
in the figures. Finally, a cutoff in the spectrum appears because of  weakly   
scattered particles escaping through the introduced upstream boundary.

\begin{figure}[t!]
\plotone{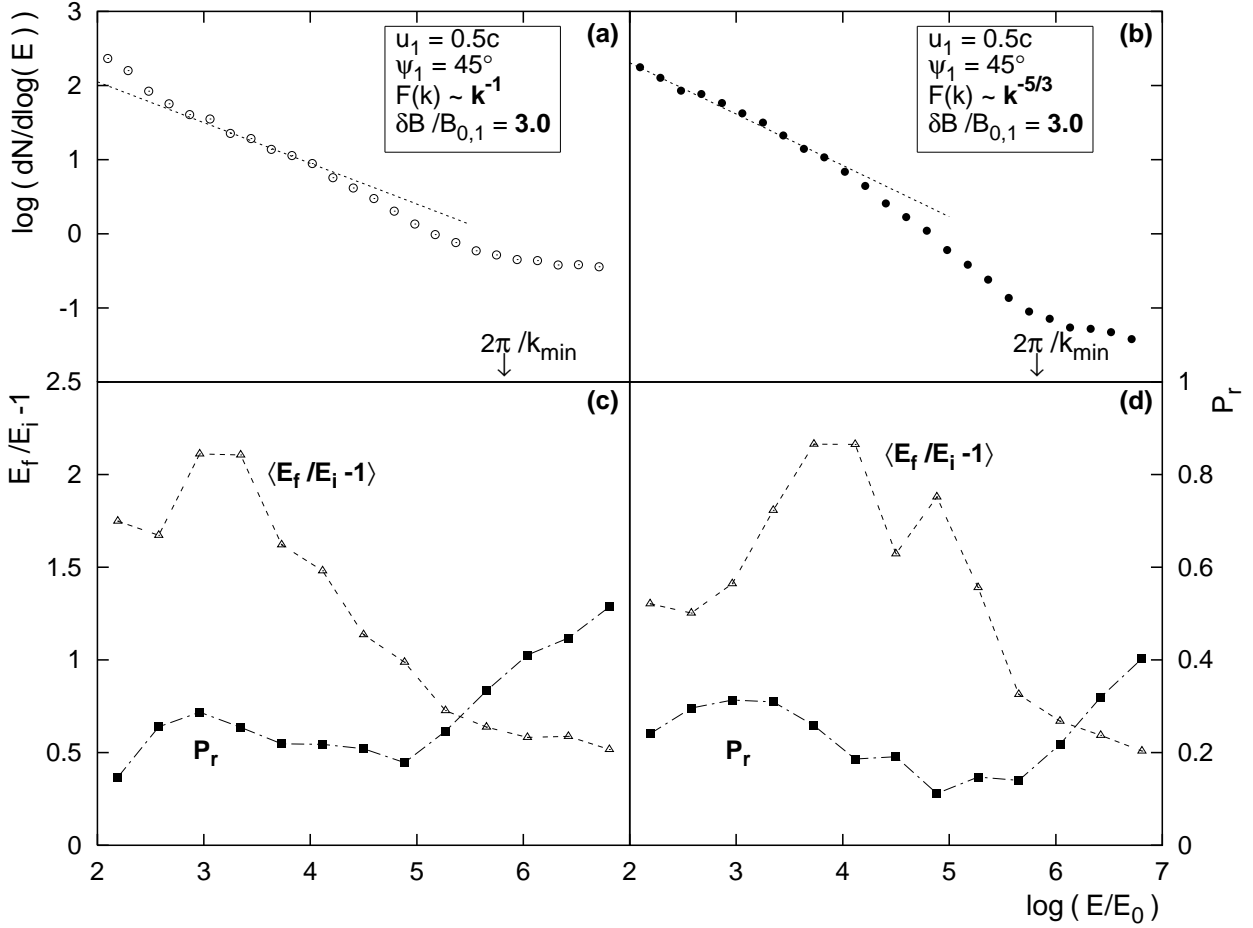}
\caption{\small Probability of particle reflection $P_r$ and 
the mean 
energy gain of reflected particles $\langle E_f/E_i-1\rangle$ in the energy
range where the particle spectrum deviates from a power-law form. The
respective parts of particle spectra are shown in the top panels. 
Power-law fits are the same as shown in Fig. 2. The values of
$P_r$ and $\langle E_f/E_i-1\rangle$ were obtained in additional simulations, 
with the 
particle injection energy $E=10$, so that the first few points are influenced by
the initial conditions. In this figure,
the energies $E_i$ and $E_f$ are measured in the upstream plasma rest frame and the
particle spectra are also calculated in this frame. To limit statistical
fluctuations, the logarithmic energy bin used in the bottom
panels is twice as large as the respective bins in the top panels. The
respective upstream-downstream transmission probability $P_{12}=1-P_r$. }
\end{figure}

Effects of the finite wavevector range can  also be observed for particles 
within the resonance range, $k_{min}\la k_{res}\la k_{max}$, for the
magnetic field perturbations considered. The change of the scattering 
conditions with the increasing particle energies should lead to the modest but
continuous variation of the particle spectral slope. The effect is visible
(Fig. 3) for the case of a
highly turbulent magnetic field $\delta B / B_{0,1}=3.0$. The noticeable
steepening of the spectrum for particle energies $\log(E/E_0)\ga 4$ 
is accompanied by a slow decrease of the probability of reflection, $P_r$, and 
the mean energy gain at an individual particle-shock interaction (Figs 3c and 3d;
for the probability of reflection and the mean energy gains calculations, see 
\S 3.4).
At higher energies, $P_r$ starts to grow again, without any increase in 
$\langle E_f/E_i-1\rangle$, which results in a harder spectrum. Thus, the main 
factor leading to the variation of a spectrum inclination is the increased 
upstream particle transmission probability and thus the increased  escape 
probability. The decrease
in $P_r$ and in the mean energy gain of reflected particles stems possibly from
the fact that for high-energy particles (but still with 
$k_{res}\ga k_{min}$) there are 
gradually fewer large-amplitude long-wave magnetic field
perturbations for an increasing particle energy and, therefore, the effective 
mean inclination
of the magnetic field at the shock decreases, leading to the effects discussed. 
These effects are not visible in the spectra for smaller $\delta B / B_{0,1}$
since the amplitudes of long-wave perturbations are moderate in these cases.
In the case of the Kolmogorov wave power spectrum yet other effects might
be
responsible for the decrease in the reflection probability. The first factor is
related to the increasing amplitude of the resonant perturbations with the
increasing particle energy, leading to a gradual breakdown of
approximate magnetic moment conservation. At the same time, the
perturbation amplitude growth leads to additional isotropization of particles
hitting the shock from upstream. This increases the mean inclination of particle
momenta crossing the shock, increasing the mean energy gain of a few reflected
particles and increases the particle upstream-downstream transmission 
probability (see Ostrowski 2002). The effect can  be also visible for particles 
forming the 
power-law part of the spectrum. As Figures 3b and 3d show,  the mean energy gain 
grows with energy for these particles, the feature not observed with the
flat wave power spectrum (Figs. 3a and 3c).

Example particle angular distributions at the shock are presented in Figure 4
for particles forming the power-law part of the energy spectrum and for
particles responsible for the flatter high-energy spectral component. 
Anisotropies of the
distributions presented for the power-law parts of the particle spectra are 
larger in the case of the Kolmogorov wave power spectrum. 
Angular distributions for particles forming the harder spectral component
hardly depend on the wave power spectrum.

\subsection{Oblique Superluminal Shock Waves}
Spectra for superluminal shocks are presented in Figure 5 for a number of the 
mean magnetic field configurations, and shock velocities (Lorentz factors)
$u_1=0.5c$ 
($\gamma_1\simeq 1.2$) and $u_1=0.9c$  ($\gamma_1\simeq 2.3$). 
For the inclination angle $\psi_1=75^o$ considered, the shock wave
propagation velocity along the mean magnetic field component is 
$u_{B,1}\simeq 1.93c$ for $u_1=0.5c$ and $u_{B,1}\simeq 3.48c$ for $u_1=0.9c$.
For the smaller $\psi_1=45^o$ and the shock velocity 
$u_1=0.9c$, the projected velocity $u_{B,1}\simeq 1.27c$.  

\begin{figure}[t!]
\epsscale{0.65}
\plotone{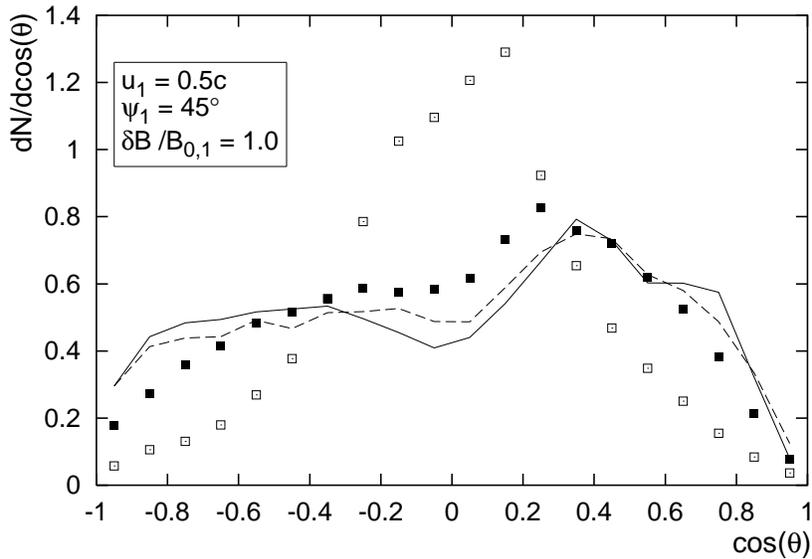}
\caption{\small Particle angular distributions at the shock front
for $\delta B / B_{0,1}=1.0$, in the shock rest frame. The angular distributions 
for particles
forming the power-law part of the energy spectrum, $1\leq\log( E/E_0 )\leq 5$,
are shown by the filled and open squares for the flat and the Kolmogorov wave 
power
spectrum, respectively. Angular distributions for particles responsible for the
harder spectral component, $5\leq \log(E/E_0 ) \leq 7$, are presented by the
solid (flat spectrum) and dashed (Kolmogorov spectrum) lines. All distributions
are normalized to the unit surface area under the respective curves. Particles
with $\cos\theta<0$ are directed upstream of the shock. }
\end{figure}

\begin{figure}
\epsscale{1.0}
\plotone{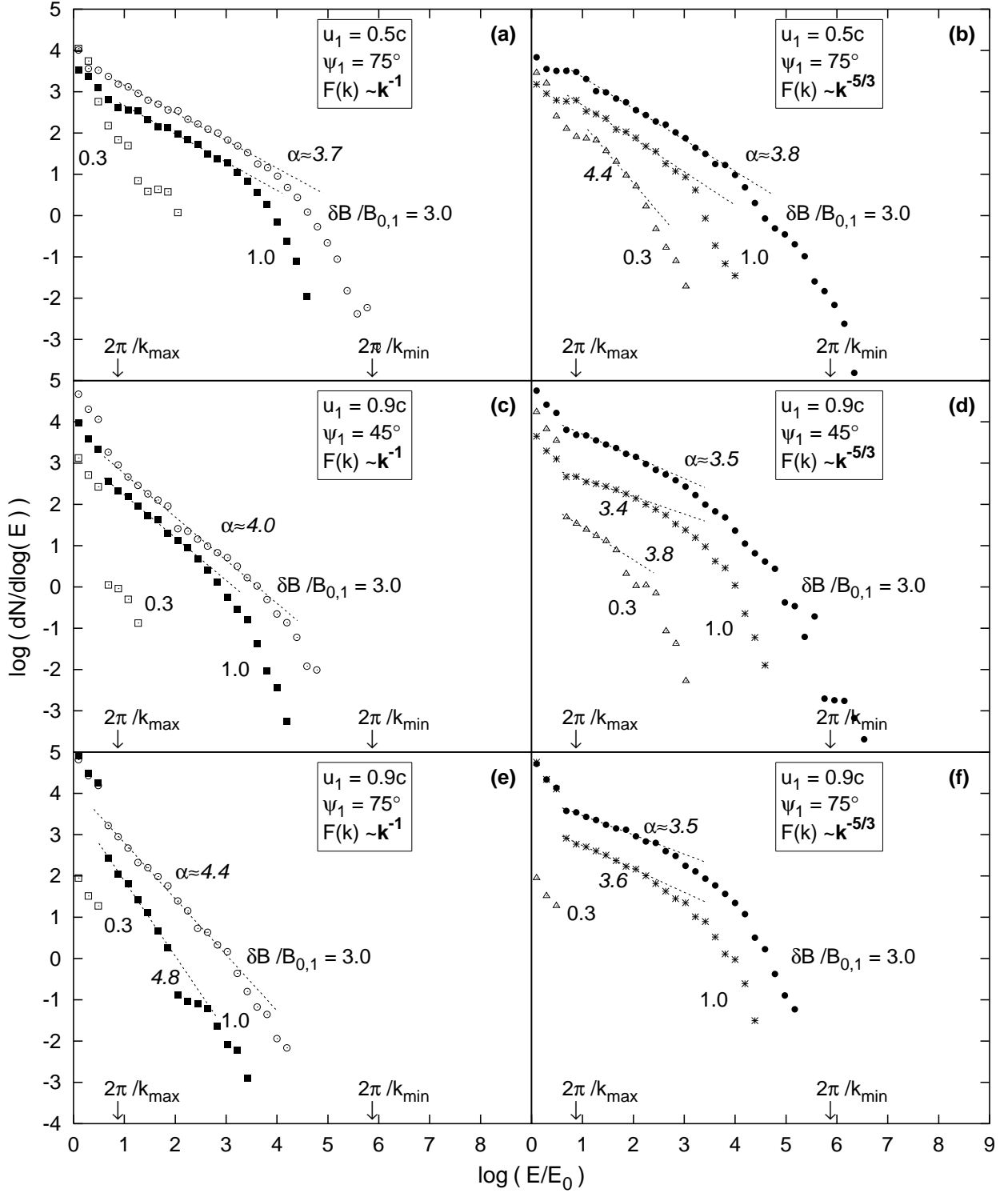}
\caption{ \small Accelerated particle spectra at oblique superluminal
shocks with $u_1=0.5c$ and $u_1=0.9c$. Parameters for each spectrum are provided
in the respective panels.}
\end{figure}

\begin{figure}[t!]
\epsscale{0.6}
\plotone{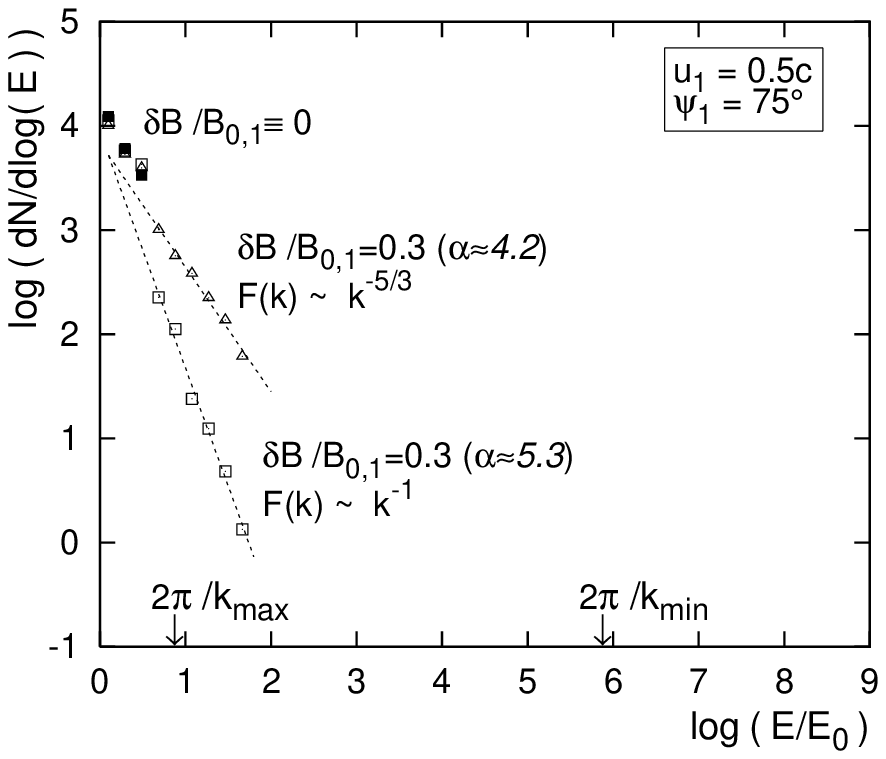}
\caption{\small Compressed particle spectra at the superluminal
shock with $u_{B,1}\simeq 1.93c$. The results for $\delta B/B_{0,1}\equiv 0$ are
presented by filled squares. The
presence of small amplitude magnetic field perturbations, 
$\delta B/B_{0,1}=0.3$,
produces power-law tails, as illustrated by the spectra obtained for 
the flat ({\it open squares}) and the Kolmogorov ({\it open triangles}) 
wave power spectra. The spectra presented are normalized in such a way as to
have the same particle weight in each spectrum. }
\end{figure}

For the small turbulence amplitude ($\delta B / B_{0,1} = 0.3$) in all cases  
studied, we approximately reproduce the results of Begelman \& Kirk (1990), showing
 a ``super-adiabatic'' compression of injected particles but hardly any or an
extremely steep power-law 
spectral tail. However, our method of particle injection does not match initial
conditions of the compression process, which takes place when the initially
isotropically distributed particles cross the shock. Because of this,
in order to directly estimate the role of this process on particle spectrum
formation, additional simulations have been performed. In these simulations, 
a set
of $N=20000$ particles with energy $E_{0,1}=E_0$ was injected $10\, r_{g,1}$ in 
front of the shock with the isotropic angular distribution  
in the upstream plasma rest frame. In these simulations we studied a
shock wave propagating with velocity $u_1=0.5c$ in the plasma with the magnetic 
field composed of the regular component only ($\delta B/B_{0,1}\equiv 0$) 
inclined at an angle $\psi_1=45^o$ to the shock normal (see Figs. 5a and 5b).
Therefore, in the absence of scattering, the particle energy gains are solely
due to the acceleration at individual shock transmissions. 
The trajectory of each injected
particle was followed until, after interacting with the shock surface, it reached 
the free escape boundary located 
 $30\, r_{g,2}$ downstream of the shock. Note
that an interaction with the
shock wave may involve a number of particle 
trajectory crossings of the shock front. The resulting particle 
spectrum is shown in Figure 6 
({\it filled squares}), where the spectra formed in the weakly 
perturbed magnetic 
field $\delta B/B_{0,1}=0.3$ are also presented for comparison. These 
spectra were calculated using the above-described method, with particle injection
upstream of the shock. One can see that the 
presence of small-amplitude
magnetic field perturbations hardly influences the particle
spectrum formation. The spectrum is essentially the
super-adiabatically compressed initial one, and the power-law part, with a sharp
cut-off at $E\sim 10^2 E_0$, is formed by a few particles only. 
The energetic tail generation 
is due to locally subluminal magnetic field configurations formed at the shock
front, which enable particle reflections or extended interaction with the shock. 
Since the locally 
subluminal field configurations are produced mainly by high-amplitude long-wave
perturbations, these processes are more efficient for the Kolmogorov-type field
perturbations. Therefore, the power-law part of the particle spectrum is
flatter in this case, when compared to the spectrum for the flat wave power
spectrum. Modifications of the compression process due to the presence of the
magnetic field perturbations depend on the shock velocity and on the mean field
configuration, as one can see from Figure 5 (see also Ostrowski 1993). For lower
$u_{B,1}>1$, the weakly perturbed magnetic field can modify the spectra, 
whereas for highly superluminal configurations the power-law part is not
generated for $\delta B/B_{0,1}=0.3$ (Figs. 5e and 5f). At larger turbulence 
amplitudes power-law sections in the spectra are again produced in the limited 
energy ranges. The spectral indices vary with varying conditions at the shock, 
including the shock velocity, the mean field inclination, and the amplitude and  
spectrum
of magnetic field perturbations. The steepening and the cutoff  occur at 
low energies, in the resonance range, contrary to the subluminal shocks at the
same turbulence amplitudes.

\begin{figure}[t!]
\epsscale{1.0}
\plotone{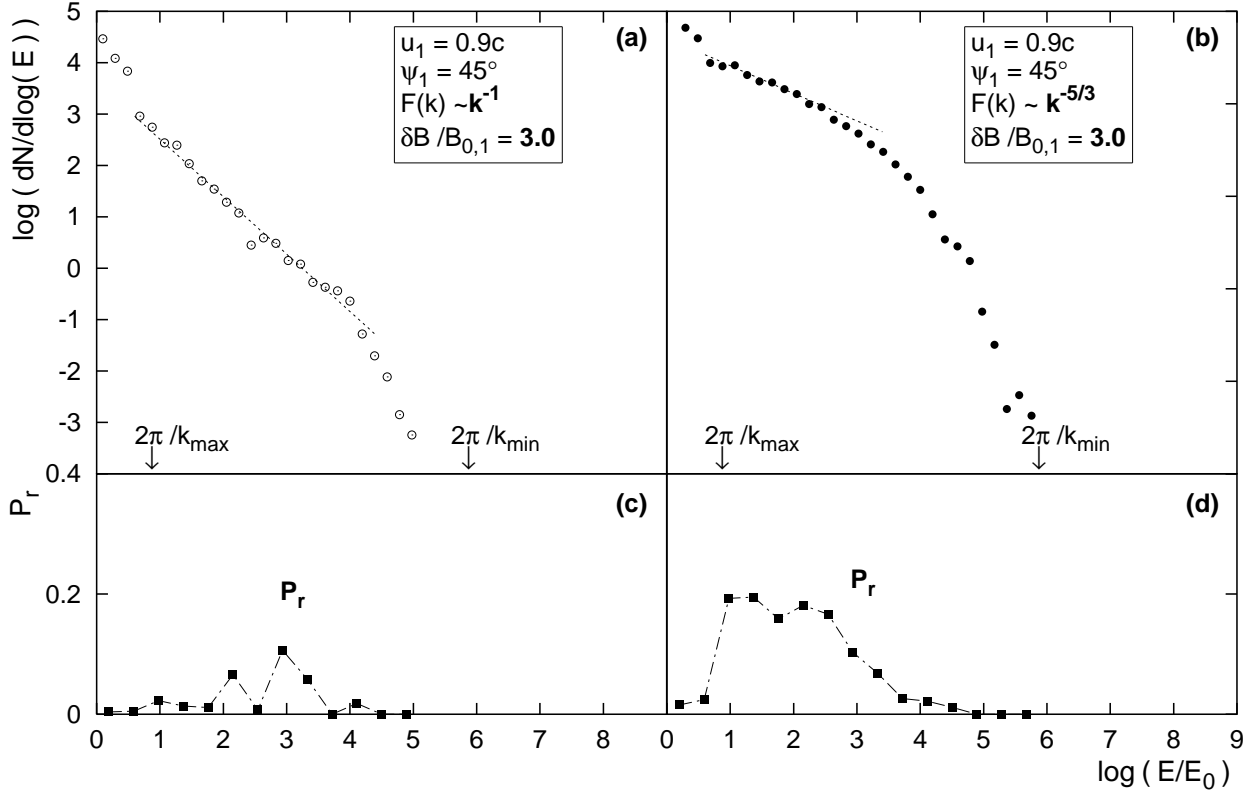}
\caption{\small Particle spectra at the superluminal shock wave with parameters
given in the panels. In the bottom panels the respective
particle reflection probability is provided. The spectra presented  
correspond to
those shown in Figs. 5c and 5d for the case of $\delta B/B_{0,1}=3.0$. }
\end{figure}

\begin{figure}[t!]
\epsscale{0.65}
\plotone{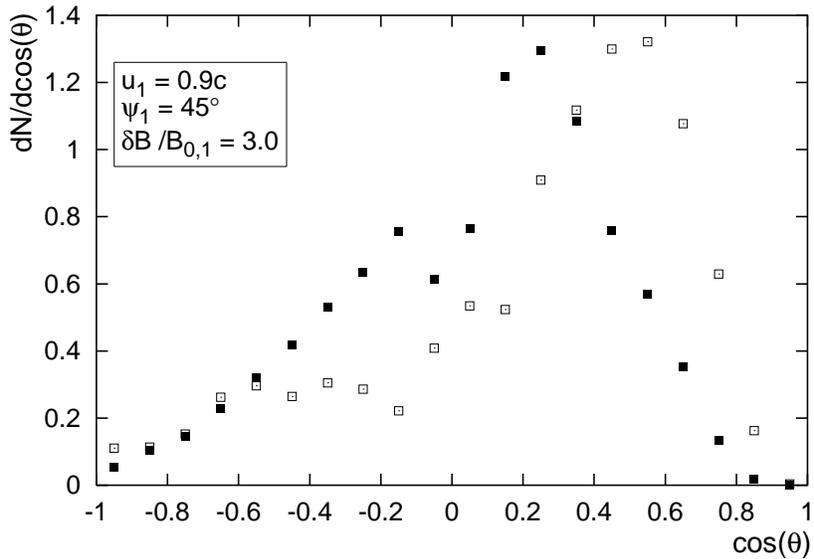}
\caption{\small Particle angular distributions at the superluminal 
shock wave for $\delta B / B_{0,1}=3.0$ in the shock rest frame. 
The angular distributions are formed by particles with energies in the range
$1\leq \log( E/E_0 )\leq 5$ and are shown by the filled and open squares for the
Kolmogorov and the flat wave power spectrum, respectively. }
\end{figure}

The spectrum steepening and the cutoff in the resonance energy range are
related to the character of the first-order Fermi process at superluminal
shocks,
in  conditions of a highly perturbed magnetic field near the shock. The
presence of three-dimensional, finite-amplitude magnetic field perturbations 
enables
efficient particle cross-field diffusion (e.g., Giacalone \& Jokipii 1994) 
but also leads to the formation of locally oblique, subluminal 
field configurations at the shock front. Such configurations are 
formed by long-wave high-amplitude magnetic field perturbations and enable
particle interactions with the shock in locally subluminal conditions. In 
some cases reflections from compressed downstream fields can also occur. 
In effect,
the energy gains of particles interacting with the shock in the locally
subluminal conditions can be much larger than those of particles interacting within the 
locally
superluminal field configurations.  The role of this factor in forming particle
spectra was originally discussed by Ostrowski (1993). For sufficiently high
turbulence amplitude, he obtained power-law spectra in a wide energy range 
because of similar scattering conditions at each particle energy. 
In our present simulations, because of 
the limited wavevector range, the scattering conditions
vary with a particle energy, leading to spectral features that have not been
discussed so far. Figure 7 shows a variation of the particle reflection
probability with the particle energy for the shock wave with $u_1=0.9c$
and  $\psi_1=45^o$. We use this parameter to describe subluminal features of the
acceleration process. Note, however, that it is not only reflections that are
responsible for the observed changes in the acceleration process; variations
in the mean energy gain at the particle-shock interaction and in the downstream
particle escape rate can also play a role. As shown in Figure 7, in the energy 
range where a
power-law part of the spectrum forms, the reflection probability changes 
slightly,
since particle resonance wavevectors are much larger than $k_{min}$. Since at a
given $\delta B/ B_{0,1}$ the amplitudes of long-wave magnetic field 
perturbations are 
larger for the Kolmogorov turbulence, $P_r$  is greater in this case. When the
reflection probability starts to decrease, the particle spectrum steepens and,
finally, for $P_r\simeq 0$, the cutoff occurs. At least two factors can be
responsible for the decrease of $P_r$. The first one is related to the limited
dynamic range of the field perturbations: with increasing particle energy 
there are gradually fewer  long waves and most of the particles meet 
effectively superluminal conditions at the shock. The second
factor is due to the accompanying increase of the resonant perturbation amplitude
for the Kolmogorov spectrum, as discussed in \S 3.1.

Examples of particle angular distributions derived at the superluminal shock 
are presented 
in Figure 8 for the case of the highly perturbed magnetic field  
$\delta B/B_{0,1}=3.0$.

\begin{figure}[t!]
\epsscale{1.0}
\plotone{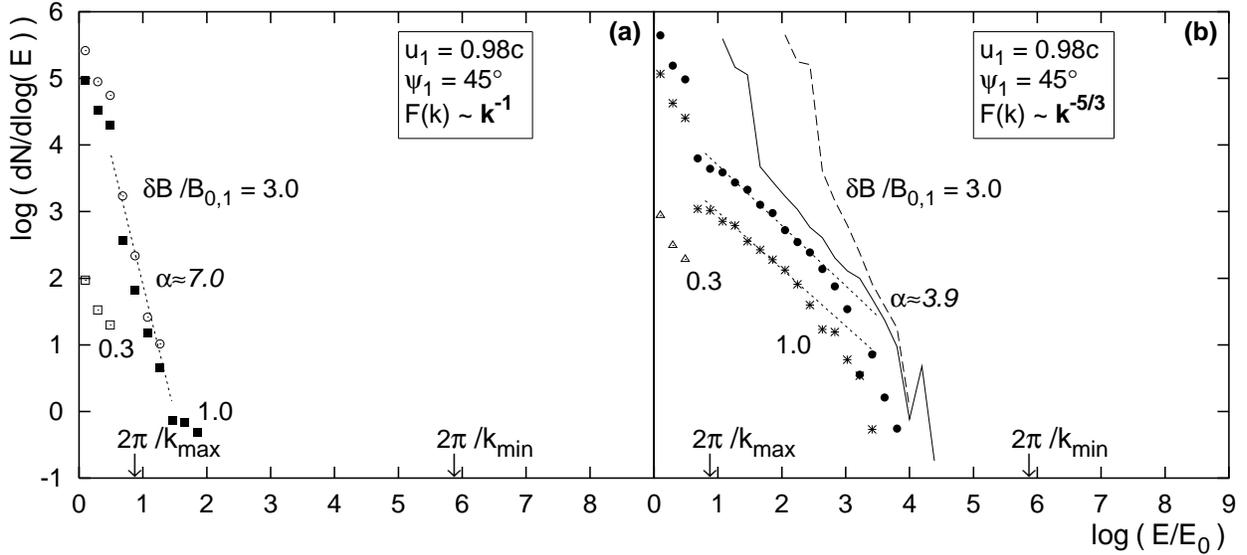}
\caption{\small Accelerated particle spectra at the
ultrarelativistic shock wave ($u_1=0.98c, \gamma_1\simeq 5$). In panel (b) the 
particle spectra obtained for higher particle injection energies, 
$E_{inj}$, are presented 
by the solid and the dashed line for $\log(E_{inj}/E_0)=1$ and 
$\log(E_{inj}/E_0)=2$ ($\delta B/B_{0,1}=3.0$), respectively. }
\end{figure}

\subsection{High $\gamma$ Shock Waves}
Particle spectra at the shock wave propagating with velocity $u_1 = 0.98c$ 
into the magnetic field with the mean field  inclined at the angle
$\psi_1=45^o$ are presented in Figure 9. The shock Lorentz factor is
$\gamma_1\simeq 5$, so one can consider the shock an 
ultrarelativistic one. As can be seen in the figure,  the main 
particle acceleration process in this case is the particle compression
at the shock: the power-law tail is formed by only a small fraction of
injected particles in the
case of a highly perturbed magnetic field. For the flat wave power
spectrum the power-law part is very steep, $\alpha\approx 7$; 
for the Kolmogorov turbulence $\alpha\approx 3.9$ with accuracy 
$\Delta\alpha\sim 0.1$. 
The spectrum cutoff occurs well into the resonance energy
range, at lower energies in comparison to the lower $\gamma$ shocks. 
This feature is independent of the initial energy of particles injected at the 
shock, as illustrated by the particle spectra calculated for the other injection
energies.
\footnote{The average magnetic field strength $\langle(${\bf\em B}$_0 + 
\delta${\bf\em B}$)^2\rangle^{1/2}$
downstream of the shock is $\langle B_2\rangle\approx 30 B_{0,1}$ and the wave 
vectors
of magnitude $k_{max}$ are the resonance ones for the particles of energies 
$E\approx 2\pi\langle B_2\rangle/k_{max}\approx 18.8$. Simulations for the
other injection energies have been performed in order to check whether the cutoff
occurring in the particle spectra is not due to the particle injection
procedure applied. For the particle injection energy $\log(E/E_0)=2$ this 
procedure gives the respective wavevectors 
$k_{res}=2\pi/r_g(B=\langle B_2\rangle)\la k_{max}$ (see \S 2.2). } 
It seems
therefore that it is the turbulent magnetic field structure that becomes
effectively perpendicular and does not allow
for particle acceleration to higher energies. Note however, that the 
$\gamma_1\simeq 5$ case we analyzed was at the limit of the application range 
of our simulation method. Thus this attempt at studying the first-order Fermi
process at ultrarelativistic shock waves should be treated with some caution.

\subsection{Parallel Shock Waves}
As discussed by Ostrowski (1988b)  for nonrelativistic shocks, the presence of
finite-amplitude magnetic field perturbations modifies the character of diffusive 
particle acceleration at shock waves with the mean 
field parallel to the shock normal. The effect arises because of locally oblique 
field configurations, formed by long wave perturbations at the shock front and 
the respective magnetic field compressions downstream of the shock. As a result, 
the mean particle energy
gains can increase and the particles reflected from the shock front can occur. 
The analogous phenomena should occur at relativistic shocks 
(see Ostrowski 1993).

In the numerical simulations of Bednarz \& Ostrowski (1996)
of the first-order Fermi acceleration at parallel
mildly relativistic shocks, the acceleration time scale reduces with increasing 
turbulence level, but no spectral index variation is observed. The same behavior
occurs for parallel ultrarelativistic shocks, where the value of the spectral
index for a given shock velocity is independent of the magnetic field
perturbation amplitude (Bednarz \& Ostrowski 1998). However, the considered 
acceleration models apply very 
simple modeling of the perturbed magnetic field effects by introducing particle
pitch-angle scattering. Such an approach does not allow for studying the effects of
long-wave perturbations. We therefore decided to investigate such processes with
the ``realistic'' magnetic field turbulence model applied in the present 
simulations.

\begin{figure}[t!]
\epsscale{1.0}
\plotone{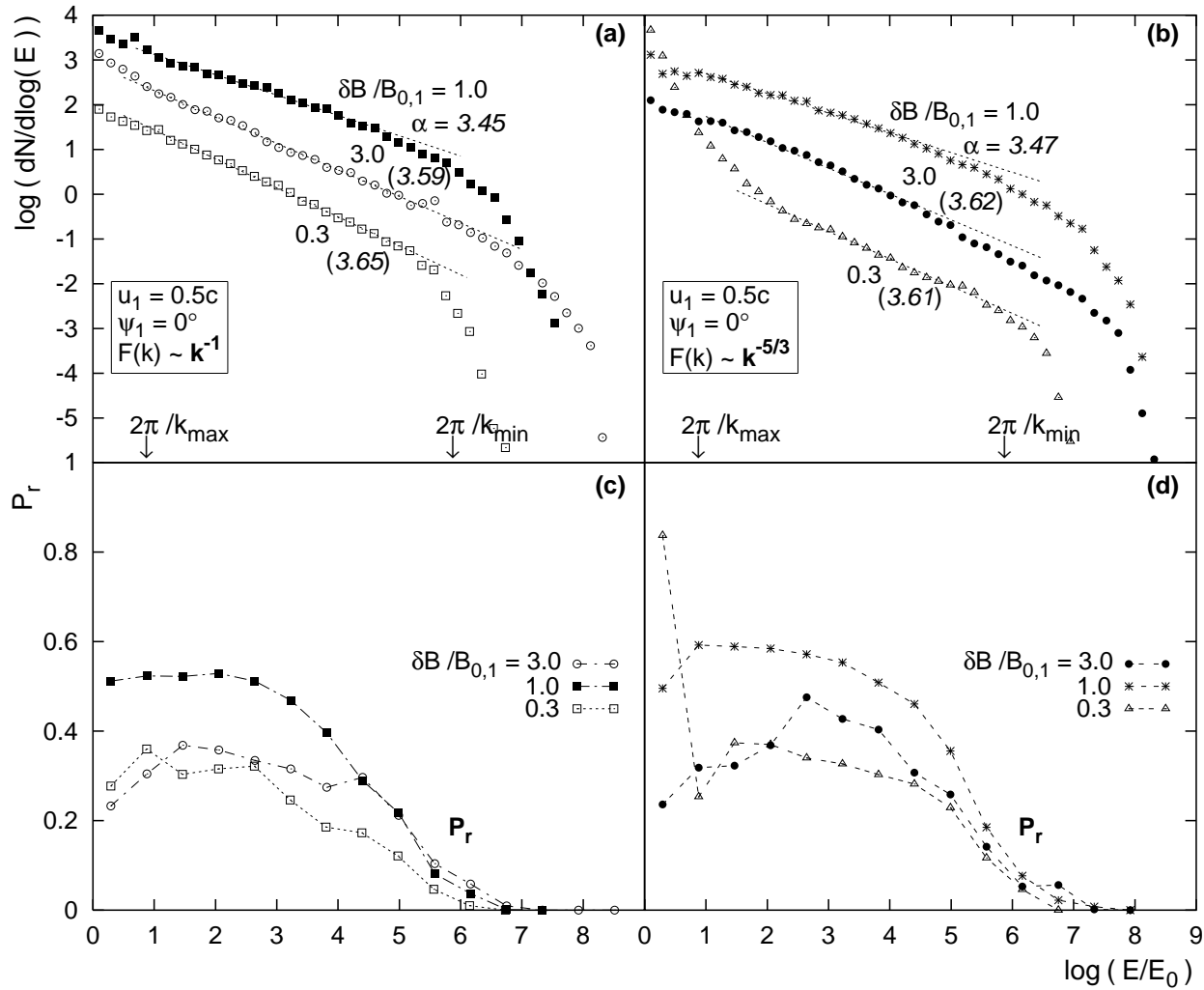}
\caption{\small Accelerated particle spectra at the parallel shock 
wave with $u_1=0.5c$. For upstream particles, the
probability of reflection from the shock, $P_r$, is presented as a function 
of a particle energy for the respective particle spectra above. }
\end{figure}

\begin{deluxetable}{ccc}
\tablecaption{Spectral indices obtained for parallel shocks with $u_1 = 0.5c$}
\tablewidth{0pt}
\tablehead{
$\delta B / B_{0,1}$ & \multicolumn{2}{c}{Spectral Index $\alpha$}\\
 & \colhead{$q=0$} & \colhead{$q=5/3$}}
\startdata
$\ll 1$ (\small analytic)\tablenotemark{a} & 3.72 & 3.74 \\
0.3 & 3.65 & 3.61 \\
1.0 & 3.45 & 3.47 \\
3.0 & 3.59 & 3.62 
\enddata
\tablenotetext{a}{Values of $\alpha$ calculated by 
Heavens \& Drury (1988) for small amplitude perturbations, 
$\delta B / B_{0,1}\ll 1$.}
\end{deluxetable}

In Figure 10 we present particle spectra for the mildly relativistic parallel 
shock wave with  $u_1 = 0.5c$. The spectral indices fitted to  the power-law 
parts of the spectra are listed  in Table 1.
One can note that the particle spectral indices 
deviate from the small-amplitude results of the pitch-angle diffusion model; 
the 
values of $\alpha$ calculated by Heavens \& Drury (1988) are also presented in
the table. In addition, increasing magnetic field perturbations produce 
nonmonotonic changes in the particle spectral index, the feature that has 
not been discussed for parallel shocks so far. 
Analogously to oblique shock waves, our particle spectra  are non--power-law ones 
in a full energy range, and the shape of the spectrum varies with the amplitude 
of the turbulence and the wave power spectrum.

The nonmonotonic variation of the spectral index with the turbulence amplitude
results from modifications of the particle acceleration process at the shock.
The long-wave finite-amplitude perturbations produce locally oblique
magnetic field configurations and lead to the occurrence of, e.g., particles 
reflected from
the compressed field downstream of the shock. In order to quantitatively
estimate effects of such locally oblique field configurations on the particle 
spectrum formation, we calculated the mean reflection rate and the mean energy 
gains of particles interacting with the shock.
We used the following method (see Ostrowski 1988b). The process of an upstream 
particle interaction with the shock begins when it crosses the shock front 
downstream for the first time. A given particle is assumed to continue its 
interaction with the shock if, after crossing the shock, it reaches the shock 
front again not later than $1.1 t_{g, max}$ and, at the same time, its distance 
from the shock does not exceed 2$r_{g, max}$. The particle gyration time 
$t_{g, max}$ and the gyroradius 
$r_{g, max}$ are derived for the average magnetic field $\langle(${\bf\em B}$_0 + 
\delta${\bf\em B}$)^2\rangle^{1/2}$ in the respective local plasma rest frame. 
A particle trajectory is integrated, including possible successive shock
crossings, until it exceeds one of the temporal or spatial barriers thus defined.
When it takes place downstream of the shock, the particle is counted as the 
transmitted one. If it occurs upstream, the particle is assumed to be the 
reflected one. A reflected particle starts its new interaction when it 
encounters the shock front again. A transmitted particle, after moving in the 
perturbed magnetic field, is sometime able to recross the shock again upstream 
and to start its new interaction with the shock. 
For each reflected or transmitted particle, its weight 
$w$,  final energy $E_f$, and initial energy $E_i$ are measured,
the last quantity derived in the upstream plasma rest frame at the first
particle shock crossing. In order to calculate the reflection probability $P_r$
and 
transmission probability $P_{12}$, a particle weight is added to the respective
logarithmic energy $E_i$ bin for the reflected particles ($S_r$) and the
transmitted ones ($S_{12}$). Then the reflection probability is calculated as
$P_r= S_r/(S_r+S_{12})$ and, analogously, $P_{12}= S_{12}/(S_r+S_{12})$. The
particle energy gains are derived as $(E_f-E_i)/E_i$, $E_f$ for the transmitted
particles being calculated in the downstream plasma rest frame (see Ostrowski
1988a). 

Reflection probability as a function of particle energy is presented
in Figures 10c and 10d for the respective particle spectra shown in the top panels. As
one can see, the probability of reflection depends on the turbulence amplitude 
and the amount of
field perturbations with wavelengths larger than the resonance wavelength for a
given particle. For $\delta B / B_{0,1}=1.0$ the reflection probability is
higher as compared to the other  perturbation amplitudes considered, and the
resulting
particle spectrum is flatter. For our chosen  $\delta B/B_{0,1}=0.3$ and 
$3.0$, the values of  $P_r$ (and $P_{12}=1-P_r$) do not differ considerably, 
which results in similar particle spectral indices. One can see that the
spectra obtained for the Kolmogorov case seem to exhibit a continuous slow 
steepening. Thus, the fitted power laws depend to some extent on the 
energy range chosen for the fit, with accompanying variations of $\alpha$
reaching $\Delta\alpha\sim 0.04$. One can also see a 
steep part of the spectrum at low energies for $\delta B / B_{0,1}=0.3$ in
Figure 10b. This is due to weak resonant scattering of low-energy particles, escaping 
thus along the magnetic field lines through the upstream
free escape boundary, located 30000 $r_g$ from the shock. 

\begin{figure}[t!]
\epsscale{0.65}
\plotone{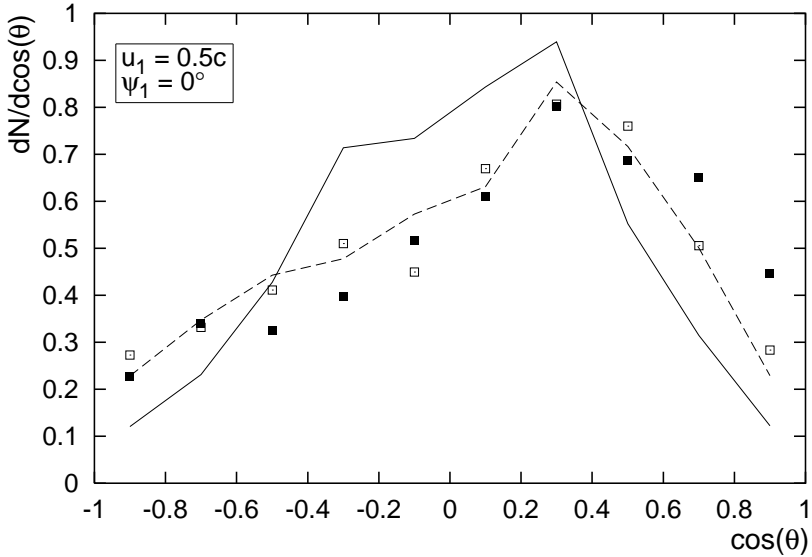}
\caption{\small Particle angular distributions at the parallel 
shock wave with $u_1=0.5c$ formed by particles with energies in the range
$1\leq\log( E/E_0 )\leq 5$ for the small ($\delta B/B_{0,1}=0.3$; {\it squares}) and
large ($\delta B/B_{0,1}=3.0$; {\it lines}) turbulence amplitude. The open squares and
dashed line are for $F(k)\sim k^{-1}$; the filled squares and solid line are for 
$F(k)\sim k^{-5/3}$. }
\end{figure}

The presented reflection probabilities decrease  
at high particle energies because of the limited dynamic range of the magnetic 
field turbulence. The locally oblique field configurations are mainly formed by 
long wave perturbations ($k \ll k_{res}$; see Ostrowski 1988b). For high-energy
particles with
$k_{res} < k_{min}$ there are no such long waves and the upstream 
particles can be only transmitted downstream of the shock. In these conditions,
the acceleration process should converge to the ``classic'' parallel shock 
acceleration model, but in our simulations particles can move far to the 
introduced escape boundary  at $x_{max}=120 r_g(E)$, forming a cutoff.

Figure 11 shows particle angular distributions for particles forming  power-law
parts of the energy spectra. For the small amplitude of magnetic field 
perturbations, 
$\delta B/B_{0,1}=0.3$, the derived distributions approximately correspond to
those obtained by Heavens \& Drury (1988). The presence of large-amplitude
perturbations ($\delta B/B_{0,1}=3.0$) has a noticeable effect on the angular
distributions, in comparison to the $\delta B/B_{0,1}=0.3$ case, only  for the
Kolmogorov turbulence spectrum.

\section{SUMMARY AND FINAL REMARKS}
The present work is intended to study  the first-order
Fermi acceleration processes acting at relativistic shocks. To derive the
stationary particle spectra we apply the
test particle approach and neglect the possibility of radiative losses or of
the second-order acceleration process acting in the turbulent medium
near the shock. The hybrid method applied for a particle trajectory integration
allows us to include essential characteristics of  particle motion important
for the acceleration process. In comparison to  
previous work, we include a few ``realistic'' features of the magnetic field
turbulence. The
field perturbations are imposed on the homogenous component, which allows us 
to analyze conditions with different mean magnetic field inclinations to the
shock normal and a range of turbulence amplitudes.
We use an analytic model for the perturbations, forming the 
power-law spectrum within a wide wavevector range. The turbulent 
magnetic field is continuous across the shock; the downstream field is 
derived from the upstream one with the respective shock jump conditions. 

The modeling shows how the resulting spectra of accelerated particles depend
on the shock velocity and the turbulent magnetic field structure considered. 
In particular, we
demonstrate the effects of the finite wavevector range of the turbulence, leading
to deviations of the derived spectra from the usually considered power-law form
(e.g., a harder spectral component occurring at high particle energies in 
subluminal shocks, the spectrum steepening and the formation of the cutoff in 
the resonance energy range in
superluminal shocks). Furthermore, we discuss the role of the long-wave magnetic
field perturbations in forming particle spectra. We show that the formation of 
locally subluminal configurations is important for a power-law spectrum 
generation at superluminal shocks, but long-wave field perturbations lead to 
significant modifications of the diffusive acceleration process at parallel
shocks as well. In our modeling, we 
reproduce also a number of previously obtained results. In particular, we
demonstrate that particle spectral indices depend on the mean magnetic field 
configuration and on the amplitude and the wave 
power spectrum  of field perturbations. 
A strong dependence of test particle spectra on conditions at 
relativistic shock fronts means that reliable theoretical modeling of real 
astrophysical sources in which such waves occur can be difficult. 
 
The first-order Fermi process can be applied to sufficiently energetic
particles only, which have their gyroradii much larger than the dissipative
thickness of the shock, defined possibly by gyroradii of thermal ions
present in the plasma. The results of the present paper are valid for such 
high-energy particles.
Particles of lower energies are expected to be accelerated by different 
processes, such as those discussed by, e.g., Hoshino et al.
(1992) or Pohl et al. (2002).

In the present study the particle radiation and other energy 
losses have been neglected. The role of losses in each particular object
can be estimated from comparison of particle acceleration timescales at the 
shock and the respective timescales for radiation losses downstream of the 
shock. Since at relativistic shock waves the particle acceleration time-scale 
depends strongly on the shock parameters \citep{bed96} and the background 
conditions can vary substantially between various objects, such an analysis is
not straightforward and requires an independent simulations for each object
studied. 
Qualitatively, if the loss mechanism is a synchrotron radiation or inverse 
Compton scattering, the cutoff in the spectrum at the shock should appear at 
high particle energy \citep[see][]{kir87b}. Downstream of the shock, such a 
spectrum evolves with the distance from the shock front and a proper modeling of 
the spectrum requires again the detailed knowledge of the physical parameters 
near the shock.

Further progress in the study of relativistic shock particle 
acceleration is impossible without understanding of the microphysics of 
relativistic shocks. A noticeable advance in this field may result from
application of particle-in-cell simulations, which allow for consistent
studies of the shock structure, particle injection  and  magnetic
field generation at the shock (e.g., Drury et al. 2001; Schmitz et al. 2002; 
Nishikawa
et al. 2003; Frederiksen et al. 2003a,b). Realistic modeling of particle
acceleration at relativistic astrophysical shocks, which have to incorporate 
results of such studies, requires a full plasma nonlinear description. This
should take into account appropriate boundary conditions, the second-order
acceleration processes, the accelerated
particle influence on the shock wave structure (including the possibility of the
existence of a cosmic-ray precursor), and the magnetic field turbulence 
structure in the
vicinity of the shock. Monte Carlo numerical simulations have recently been
used to study nonlinear particle acceleration at shocks modified by the
back-reaction of accelerated particles (Ellison \& Double 2002). Estimates of the
effectiveness of second-order Fermi process acting in a strongly turbulent 
field downstream of the shock made by Dermer (2001) suggest that this process 
is able to produce high-energy particles. To make such models more realistic,
there is a need for detailed studies of the problem of magnetic field 
generation and turbulence evolution near/at the shock.  

Let us also stress that
the issue of MHD turbulence generation at the shock is of great importance
for ultrarelativistic shocks. Such shock waves are suggested to be
the gamma-ray burst sources and may also produce ultra--high-energy
cosmic rays. Modeling of the burst afterglows spectra often yields
results pointing to the asymptotic spectral index $\alpha\approx 4.3$ 
($\sigma\approx 2.3$) for
radiating electrons, which is sometimes interpreted as observational
confirmation of the correctness of theoretical models proposed for 
ultrarelativistic shock acceleration. As mentioned in \S 1, these models
consider highly turbulent conditions near the shock. Standard hydrodynamic
gamma-ray burst models assume the magnetic field to be generated locally at the
ultrarelativistic shock \citep[e.g.,][]{med99,nis03,fre03a,fre03b}, but the 
effectiveness of the generation mechanism and the resulting perturbation 
spectrum is uncertain \citep[but see][]{fre03b}. 
If this effectiveness is low, the
particle spectra produced at high-$\gamma$ shocks  are
expected to be much steeper than the asymptotic ones (see Bednarz \& Ostrowski 1998
and also the results of our simulations for $u_1=0.98c$), suggesting possibly a
different mechanism for particle acceleration \citep[e.g., simplified modeling
of electron acceleration by][and 
later particle-in-cell simulations]{hos92}. 
Therefore,  magnetic
field turbulence generation problems have to be analyzed in detail to make
realistic estimates of the importance of the first-order Fermi process on the
observed parameters of gamma-ray bursts.
Generation of ultra--high-energy cosmic rays at such shocks requires the
respective long-wave field perturbations too. It is highly uncertain whwther such
processes take place.

We have performed our modeling in conditions defined in units of a particle
gyroradius and a mean inclination of the magnetic field. The respective scaling
of these conditions by choosing values for $B_{0,1}$, $\psi_1$, and the
turbulence spectrum allows us, at least formally, to apply these results to
various astrophysical objects, as long as losses are unimportant. For situations
with radiative processes playing a role, one should repeat the simulations for given
physical values of the model parameters and the respective radiative-loss terms.

\acknowledgments

The work was supported by the Polish State 
Committee for Scientific Research in 2002--2004 as research project 2 P03D 008
23 (J. N.) and in 2002--2005 as research project 
PBZ-KBN-054/P03/2001 (M. O.).

\appendix

\section{SHORT WAVES IN OUR HYBRID APPROACH}
The influence of short waves ($\lambda\ll r_g$)  on particle
trajectories is treated here as random, small-amplitude angular perturbations of 
particle momenta with a discrete scattering procedure 
described, e.g., by Ostrowski (1991). In this procedure, a particle momentum 
direction is perturbed at discrete instants of time by a small angle 
$\Delta\Omega$. 
Fitting of this approach to the magnetic field turbulence spectrum considered 
and varying particle energies requires
determination of the scattering probability distribution for $\Delta\Omega$.
This is determined by the additional simulations described below.

Short waves are defined as the ones with 
$\lambda < \lambda_{min}=0.05 r_{g}(B=B_{0,1})$.
In analogy to the approach described by Ostrowski (1991), in our
simulations, particle trajectory perturbations are introduced along the
``exact'' 
orbit in the magnetic field composed of the mean field and turbulent 
fluctuations with wavelengths larger than $\lambda_{min}$. 
Particle momentum direction perturbations are performed at discrete instants 
of time separated by $\Delta t$. In the present simulations, 
$\Delta t = 0.01 t_g (t_g=2\pi r_{g}(B=B_{0,1})/c)$. 
At each scattering, the value of the scattering amplitude
$\Delta\Omega$ is selected at random from the scattering probability
distribution. 
Since $\Delta\Omega$ stands only for the angular distance between the original and
the scattered momentum vectors on the sphere of $|${\bf\em p}$|= const$,
one also has to specify the scattering azimuthal angle, in order to derive the
scattered vector orientation. The scattering azimuthal angle is randomly chosen 
from the range $[0,2\pi]$. 

In the additional simulations determining the scattering probability 
distribution, a large number of particles (usually $N=10^4$) were injected at 
random positions and random momentum vector orientations into the turbulent 
magnetic field composed of the short waves only. The trajectory of each particle was
subsequently integrated over the time $\Delta t$ when the scattering angle 
$\Delta\Omega$ was derived and the particle weight ($w=1$) was added to the 
respective $\log\Delta\Omega$ bin. The scattering probability
distribution obtained was ultimately normalized to unity.
Because of particle interactions with  small-scale field fluctuations, 
particle  spatial 
positions diverge from the ``exact'' trajectory in the magnetic field composed of
the mean field and long waves. The shifts introduced are small, however, and we 
neglect them 
in order to simplify the scattering model (there are no spatial scatterings
accompanying the angular ones). Since the downstream magnetic field
structure is different from the upstream one the scattering probability
distribution was determined independently in this region. 
We emphasize that the probability distribution for $\Delta\Omega$ is not 
model-assumed (see Kirk \& Schneider 1987b; Ostrowski 1991), but is determined 
by a particular pattern of short waves present in the simulations.

\begin{figure}[t!]
\epsscale{0.65}
\plotone{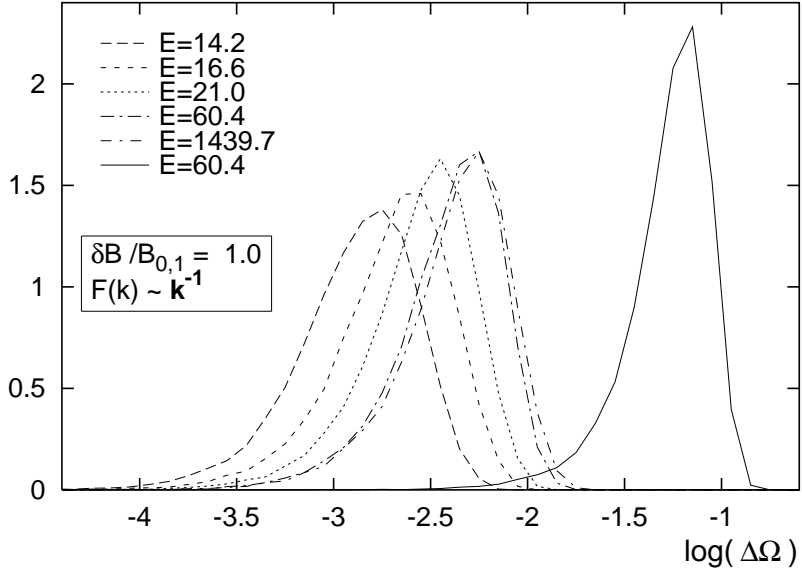}
\caption{\small Angular scattering probability distributions for selected
particle energies for the
upstream turbulence amplitude $\delta B/ B_{0,1} = 1.0$ and the flat wave power
spectrum. The solid line represents distribution of the full turbulent magnetic 
field influence on particle
motion at the same time intervals. All distributions are normalized to unity. }
\end{figure}

In the wavevector range of the magnetic field perturbations considered, the 
condition for the short wave fluctuations ($\lambda < \lambda_{min}(E)$) is 
fulfilled by different  
sets of waves for different particle energies. 
The scattering conditions and, consequently, the scattering probability 
distribution vary continuously with a particle energy.  
In our simulations, we calculate these probability
distributions for a discrete set of energies. Since our turbulent field is 
modeled by a discrete wave power spectrum, the consecutive energies are selected
in such a way that corresponding short-wave turbulent fields differ by a few 
waves only.
In effect, each subsequent distribution differs modestly from the previous one.
Example scattering probability distributions are presented in Figure 12 
for the upstream field with $\delta B/ B_{0,1} = 1.0$ and the flat wave power
spectrum. 
The mean scattering angle due to short-wave perturbations increases with a 
particle energy, but it is small in comparison with the full (including all
waves) magnetic field 
influence on particle motion, represented in the figure by the solid line
for the selected $E=60.4$.

\begin{figure}[t!]
\epsscale{0.65}
\plotone{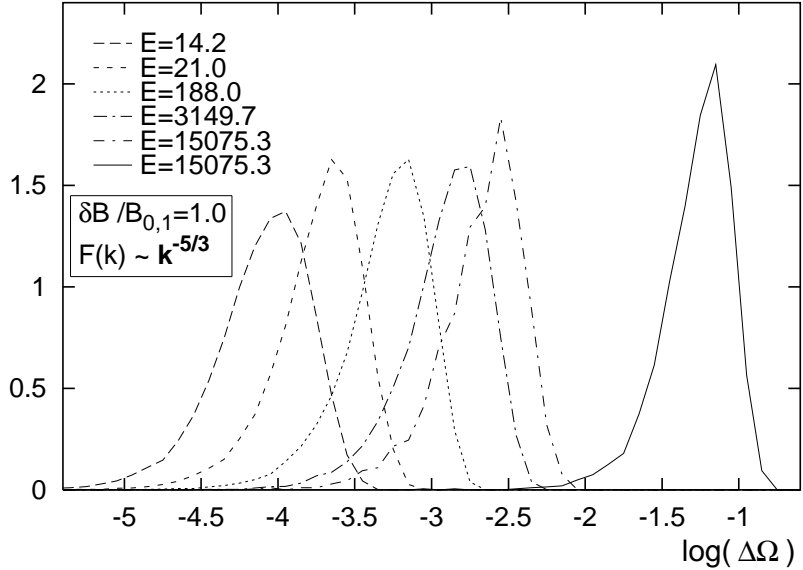}
\caption{\small Angular scattering probability distributions for the
upstream field with $\delta B/ B_{0,1} = 1.0$ and the Kolmogorov wave power
spectrum. The full magnetic field influence is represented by the solid line. }
\end{figure}

In Figure 12 the probability distributions corresponding to  highest 
energies overlap, which indicates similar scattering conditions at these
energies. In fact, the most efficient scattering is provided by the
long-wavelength part of the short waves considered at any given particle energy,
whereas the
smallest scales do not contribute significantly to trajectory perturbations.
In the turbulent magnetic field with the flat wave power spectrum, the field
energy in the waves responsible for scattering is similar for higher energy 
particles, which results in statistically identical scattering 
probability distributions in a wide energy range. This is not the case for the
Kolmogorov wave spectrum, where
the scattering conditions vary with particle energy, leading to  variations
of the scattering probability distribution, as shown in Figure 13. 
As one can see, the mean perturbation angle increases continously with 
particle energy in
this case. Finally, for the highest particle energies all the waves form 
the small-scale turbulence and the mean
perturbation angle starts to decrease.

\section{TESTS OF THE SIMULATION METHOD}
The Runge-Kutta routine we use for trajectory integration is accurate to fifth
order. The embedded fourth-order formula estimates the truncation error at a
given point of a particle orbit, which allows for an appropriate step-size 
adjustment
to ensure integration with the accuracy required. We have set the truncation 
error at
every integration step to be less than $10^{-7}$. The accuracy imposed allows
for a precise and efficient trajectory calculation. With this choice, the
 particle
energy between shock crossings is conserved to better than one part in $10^{7}$.
Integration with small but still finite accuracy produces inevitably  
numerical noise, which influences the stability of calculated trajectories.
This noise can be regarded as resulting from an additional small-scale magnetic 
field turbulence.
We have checked, via diffusion coefficient derivations (see Ostrowski 1993),
that the energy density in these fluctuations is small and does not 
change noticeably the wave power spectrum assumed for the simulations. 
Our trajectories are
stable over at least several particle gyroperiods. In the  vicinity of a
relativistic shock wave, very few particles that form the spectrum stay on 
either side of the shock front for many gyroperiods between successive shock 
crossings. The
exception may be particles of energy near the injection energy $E_0$, which only 
weakly interact with the turbulent field. However, the part of the spectrum 
formed by these particles is already influenced by the initial conditions.

In order to check whether the method of particle spectra and angular distribution 
measurement works properly, additional simulations were performed. 
An isotropic particle
source was located at some point in the plasma. On both sides of that
point a set of five parallel barriers was placed, each one several particle
gyroradii away from every other. Trajectories of $N=1000$ particles originated at
the particle source were subsequently calculated then. Every time a particle
crossed one of the barriers, a quantity $1/(|v_x| + 0.005)$ was added to the 
respective $\cos(\theta)$-bins, where $\theta$ is the angle between the particle
momentum and the normal to a given barrier. Trajectory calculations were
finished when a particle escaped far from the external barriers. The  resulting
particle angular distribution, averaged over 10 different sets of 
particles and 
realizations of the turbulent magnetic field, was isotropic, as expected. Small
departures from isotropy due to statistical errors occurred only near
$\cos\theta\approx 0$, so  angular distributions presented in \S 3 might
be measured with worse quality near that value of $\cos\theta$. These simulations
were performed for the upstream turbulent magnetic field with 
$\delta B/ B_{0,1} = 1.0$ and the flat wave power spectrum.

In our hybrid approach, used for particle trajectory 
integration, the short waves, both upstream and downstream of the shock, are 
defined by a particle gyroradius in the {\it upstream} mean magnetic field 
($\lambda < \lambda_{min}=0.05 r_{g}(B=B_{0,1})$). However, in
cases of higher turbulence amplitudes, the average magnetic field can be much
higher then $B_{0,1}$, especially downstream of the shock, where the magnetic
field is amplified by compression. Particle gyroradii are then respectively smaller than 
$r_{g}(B=B_{0,1})$. In effect, some part of the resonance wavevectors for a
particle of a given energy is treated in the simulations as a short wave
turbulence, whose influence on particle trajectories is modeled by the scattering
terms. In order to check that the definition of short waves used in our
simulations does not generate false results, additional simulations were done 
with the modified definitions of short-wave perturbations. We have checked all 
the spectra presented in the paper for the upstream turbulence amplitude 
$\delta B/ B_{0,1} = 3.0$. For this perturbation amplitude, the average
upstream magnetic field is equal to $\langle B_1\rangle\approx 2.3$, and 
the definition of short waves that we adopted is 
$\lambda < \lambda_{min}=0.02\, r_{g}(B=B_{0,1})$ in this case. The average 
magnetic field value downstream of the shock depends on the mean field
configuration and the compression factor $r$. For parallel shocks
propagating with velocity $u_1=0.5c$, $\langle B_2\rangle\approx 10$ and the
definition of short waves is $\lambda < \lambda_{min}=0.005\, r_{g}(B=B_{0,1})$.
For the above definitions of the short-wave turbulence, the scattering probability distributions have been determined. 
The particle spectra obtained in the test simulations with the modified
definition of short waves are, in the limit of statistical accuracy, 
the same as those calculated in the
simulations with the unmodified definition of small-scale perturbations.
\footnote{The value of the average downstream magnetic field can be 
larger for faster shocks (e.g. for $u_1=0.9c$, $\psi_1=45^o$, 
$\langle B_2\rangle\approx 16$), so 
the required $\lambda_{min}$ should be respectively smaller. However, for 
such small 
$\lambda_{min}$ the simulation time would be very long.}

The most important test concerns the hybrid approach we use in the simulations.
A straightforward way to check whether it gives the correct results would be to
perform simulations by integrating exact particle trajectories in the magnetic
field with all the turbulence scales present, without introducing the scattering
terms. However, it was impossible to perform such modeling with the full 
range of wavevectors considered ($k_{max}/k_{min}=10^5$) because of the 
 excessive simulation time required.
In this situation, we have performed such simulations for a limited wavevector 
range, with $k_{min} = 0.01$ and 
$k_{max} = 10$. We have investigated the acceleration process at the shock wave 
moving with velocity 
$u_1 = 0.5c$, the mean upstream magnetic field inclined at an angle 
$\psi_1 = 45^o$ to the shock normal, and the amplitude of the field turbulence 
$\delta B/ B_{0,1} = 1.0$ with the flat wave power spectrum. In addition, 
parallel simulations were performed with the use of the original hybrid
approach, with the same simulation parameters. 
Both methods yielded the same results within the statistical accuracy: the 
particle spectrum is a power-law in a finite energy range, followed by the 
harder component. Statistical accuracy of the spectral index calculation for the
modeling without scattering terms was here rather low, because these simulations 
have been performed 
for only one set of $N=100$ injected particles and one realization of the 
turbulent magnetic field.

\end{document}